\documentclass[journal=aesccq,manuscript=article,layout=twocolumn]{achemso}

\usepackage[version=3]{mhchem} 
\usepackage{amssymb}
\usepackage[dvipsnames]{xcolor}
\usepackage{booktabs}
\usepackage{siunitx}
\usepackage{ulem}

\usepackage{url}
\usepackage{rotating}
\author{Seda I\c{s}{\i}k}
\email{ishik.seda@gmail.com}
\affiliation[ITU]
{Eurasia Institute of Earth Sciences, Istanbul Technical University, 34485, Istanbul, Türkiye}
\alsoaffiliation[BoUn]
{Kandilli Observatory \& Earthquake Research Institute, Bogazici University, 34684, Istanbul, T\"urkiye}
\author{Mohit Melwani Daswani}
\email{mohit.melwani.daswani@jpl.nasa.gov}
\affiliation[JPL]{Jet Propulsion Laboratory, California Institute of Technology, Pasadena, CA 91109, USA}
\author{Emre I\c{s}{\i}k}
\affiliation[MPS]{Max-Planck-Institut f\"ur Sonnensystemforschung, Justus-von-Liebig-Weg 3, 37077 G\"ottingen, Germany}
\alsoaffiliation[TAU]
{Department of Computer Science, Turkish-German University, \c{S}ahinkaya Cd. 94, 34820, Beykoz, Istanbul, T\"urkiye}
\author{Jessica Weber}
\affiliation[JPL]{Jet Propulsion Laboratory, California Institute of Technology, Pasadena, CA 91109, USA}
\author{Nazlı Olgun Kıyak}
\affiliation[ITU]
{Eurasia Institute of Earth Sciences, Istanbul Technical University, 34485, Istanbul, Türkiye}

\title[Citric acid cycle in ocean world interiors]
  {Thermodynamic constraints on the citric acid cycle and related reactions in ocean world interiors}

\begin{document}

\begin{abstract}
Icy ocean worlds in our solar system have attracted significant interest for their astrobiological and biogeochemical potential due to the predicted presence of global subsurface liquid water oceans, the presence of organics in Enceladus and Titan, and plausible sources of chemical energy available for life therein. A difficulty in placing quantitative constraints on the occurrence and effectiveness of biogeochemical reactions favorable for life and metabolism in ocean worlds is the paucity of thermodynamic data for the relevant reactions for pressure, temperature and compositional conditions pertaining to ocean worlds, in addition to uncertainties in the estimation of such conditions. Here, we quantify the {thermodynamic} viability and energetics of various reactions {of interest} to metabolism at pressures and temperatures relevant to ocean worlds Enceladus, Europa, Titan and Ganymede, and conditions relevant to the Lost City Hydrothermal Field for comparison. Specifically, we examine the tricarboxylic acid cycle (also known as TCA, Krebs cycle, or citric acid cycle) and a plausible precursor prebiotic network of reactions leading to the TCA cycle. We use \textsf{DEWPython}, a program based on the Deep Earth Water (DEW) model (which is a high pressure and high temperature extension of the Helgeson---Kirkham---Flowers equation of state used to calculate thermodynamic properties of ions and complexes in aqueous solutions), to compute the equilibrium constants and the Gibbs free energy changes for given reactions, as a function of pressure and temperature. {Using instantaneous concentrations of inorganics and organics from terrestrial microbial experiments and those derived from the Cassini mission for Enceladus, we calculate chemical affinities of reactions in the network}. We carry out similar calculations using the SUPCRT model for lower pressures and temperatures. Together, the two models span temperatures between $0$ to 1200 $^\circ$C and pressures between 1 bar--60 kbar. 

We found that across the majority of oceanic pressure---temperature profiles, certain TCA cycle species, such as citrate and {succinate}, accumulate, while others, including {fumarate} and oxaloacetate, exhibit a diminishing trend. This observation suggests that the internal conditions of ocean worlds may not thermodynamically favor a unidirectional TCA cycle, thereby implying an additional source of energy (e.g., metabolites) to overcome energy bottlenecks. Notably, we find similar bottlenecks at the Lost City Hydrothermal Field, which is undoubtedly inhabited by organisms. In the prebiotic network, we found that pyruvate and acetate exhibit remarkable stability and accumulate in substantial quantities, thereby feeding the TCA cycle through the production of citrate. In this case the oxaloacetate bottleneck within the TCA cycle is bypassed via the prebiotic pathway. We also found that the formation of all TCA cycle species from inorganic compounds (\ce{CO2} + \ce{H2}) is, {with the exception of oxaloacetate}, highly favored throughout the geotherms of ocean worlds. Although based on largely uncertain concentrations of chemical species in ocean worlds, our non-equilibrium thermodynamic predictions are rather insensitive to changes in the activities, and may aid in the interpretation of data gathered by future missions, as compositional data will become available. Specifically, spacecraft measurements of TCA cycle species in aqueous environments that {align with or} deviate strongly from our {estimations} would {have a critical impact on the search for life in ocean worlds}.
\end{abstract}

\section{Introduction}
All known life is constantly supported and sustained by intricate networks of chemical reactions occurring within cells. Among these reactions, there are certain metabolic pathways thought to be crucial for life. The tricarboxylic acid cycle (TCA cycle, also known as the citric acid, or Krebs cycle) is considered to be one of these fundamental chemical reaction networks for aerobic organisms (and anaerobic organisms in the reverse, i.e., reductive direction), and is believed to have played a role in the origin of life on ancient Earth \cite{weber_testing_2022,Muchowska20, rauscher_hydrogen_2022, ritson_cyanosulfidic_2021, guzman_prebiotic_2009, nitschke_beating_2013}. One of the {critical} steps in cellular metabolism, the TCA cycle, is a chemical reaction chain that oxidizes activated acetic acid (acetylated coenzyme A, or Acetyl-CoA) from sugars, fatty acids, amino acids and proteins to carbon dioxide and water in oxygen-respiring organisms. Considering the thermodynamic conditions required for the emergence of life and the biochemical similarity of all extant organisms on Earth, studies suggest that the last universal common ancestor to all extant organisms (LUCA) originated in chemically diverse and energetically rich hydrothermal environments containing trace elements, transition metals, $\ce{H2}$, $\ce{CO2}$, as well as various inorganic and organic species \citep{Baross85, Russell97}. Such ancient biochemistry may have involved glycolysis and the formation of Acetyl-CoA from inorganic carbon as the first steps towards metabolism \citep{RUSSELL2004, Fuchs10}. 

How the TCA cycle takes place in hydrothermal environments is investigated by many approaches. Organic reactions may have initially used metals such as iron-containing minerals as redox catalysts in the absence of available enzymes \citep{weber_testing_2022, Muchowska20, rauscher_hydrogen_2022},  and over time, these reactions could eventually have led to the emergence of metabolic pathways that utilized enzymes\cite{Springsteen2018}. In this scenario, enzymes may not have been critically important for the initial proto-metabolic reactions of life, which motivates us to consider the thermodynamics of some biologically fundamental reactions in an abiotic environment in the absence of known enzymes {(and their biological roles along with cofactors)}, as might have occurred on Early Earth and locations in the solar system where liquid water is present. 

\begin{figure*}
    \centering
    \includegraphics[width=\linewidth]{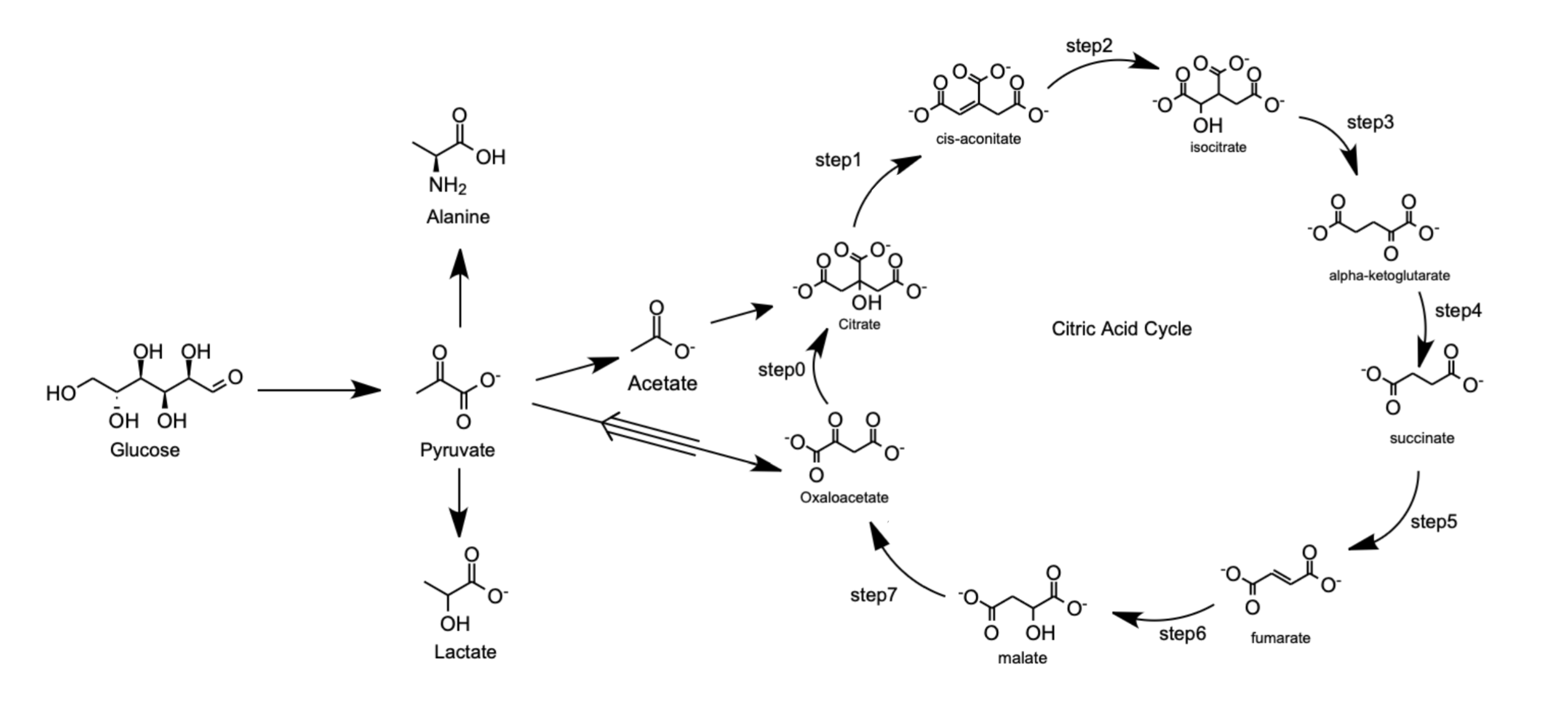}
    \caption{TCA cycle and the associated prebiotic network centered on pyruvate. The thin arrows 
    denote biologically expected pathways in typical terrestrial environments for aerobic organisms, while the double 
    arrow shows the computed {spontaneous (exergonic)} direction 
    of the associated reaction in the majority of the ocean-world interior conditions 
    considered in this study (see the Results).}
    \label{fig:network}
\end{figure*}

In addition to Acetyl-CoA, another important central species is pyruvate (\ce{CH3COCOO^-}), which is found in multiple metabolic pathways, notably in the reductive TCA cycle, as well as in pathways responsible for the synthesis of amino acids and sugars. It is thought to have been produced under hydrothermal vent conditions on Early Earth and is shown to be produced abiotically in laboratory experiments \cite{cody+00,novikov+13,stubbs+20,beyazay+23}. This suggests that pyruvate might have been an important component of a proto-metabolic network that preceded and enabled the emergence of metabolism. An example of such a network in connection with the TCA cycle is shown in Fig.~\ref{fig:network}, where glucose is converted to pyruvate via glycolysis; pyruvate can be used to form alanine, lactate, oxaloacetate, and Acetyl-CoA (or acetate). The latter two species connect pyruvate to the TCA cycle, in which pyruvate enters the initial step in the cycle as a reactant. 

In the present context, we call this proto-metabolic pathway the \emph{prebiotic network}, to emphasize its independence from the existence (or lack) of living organisms, focusing on the spontaneous occurrence of these reactions based solely on thermodynamic principles. We selected this particular network because it represents a minimal set of reactions connecting simple carbon compounds to the TCA cycle and has been proposed as a plausible precursor to more complex metabolism. We consider cofactors  such as ATP and NAD+ as chemical species participating in the reactions, rather than as part of an evolved metabolic system. This treats them as reactive molecules that could have thermodynamic equivalents in a prebiotic setting (such as polyphosphates or mineral-catalyzed redox reactions), rather than as sophisticated biological cofactors performing their modern enzymatic functions.

While we acknowledge that glucose-rich oceans are unlikely in prebiotic settings, we include this pathway not to suggest it dominated in ocean worlds, but rather to analyze the thermodynamic landscape of carbon fixation and energy coupling in these environments. The glucose-pyruvate conversion serves as a model pathway to understand how reduced carbon compounds might have been processed in early biochemical networks, even if the specific reactants and mechanisms differed from modern glycolysis. In our analysis, we consider both reference biological concentrations and more realistic trace levels (Section~\ref{sssec:som}) to understand how reaction favorability varies across different concentration regimes. Nevertheless, diverse abiogenic sugars are present in carbonaceous chondrites (e.g. \cite{furukawa_extraterrestrial_2019}), which likely represent materials similar to the building blocks of ocean worlds in our solar system.

Reactions involving the abiotic production of biologically relevant organics in hydrothermal environments at Earth's present and past ocean floors is an active field of research \cite{klingler+07,amend+11, mccollom_abiotic_2007, shock_potential_2010, shock_thermodynamics_2013, proskurowski_abiogenic_2008, song_formation_2021, mccollom_laboratory_2013}. Some of the so-called ocean worlds of the outer solar system fulfill the requirement for the presence of liquid water for these reactions, and for life, to occur. However, it is largely unknown whether the physical conditions of the interiors are conducive to such reactions \cite{camprubi_emergence_2019}. Liquid oceans are sustained underneath the icy crusts of these bodies, owing to tidal forcing from their host planet (namely Jupiter or Saturn, although Uranus and Neptune's moons may have hosted oceans in the past \cite{castillo-rogez_compositions_2023}) and radioactive and geochemical energy sources from within. 

{The prebiotic network postulated in Fig.~\ref{fig:network} uses glucose and pyruvate as starting species, though we recognize that glucose-rich environments are unlikely in prebiotic ocean worlds. Rather than suggesting glucose dominance, we include this pathway to explore the thermodynamic landscape of carbon fixation and energy transfer in potential early proto-metabolic networks, which, while being abiotic could be quite complex. Importantly, this network is not dependent on glucose: experimental work has identified multiple routes to key intermediates and reactions within the cycle\cite{Muchowska2017,Springsteen2018,Muchowska20,stubbs+20}, Our analysis considers both reference biological concentrations and the more realistic trace levels detailed in Sect.~\ref{sssec:som}, allowing us to assess reaction favorability across different concentration scenarios. The diverse abiogenic sugars identified in carbonaceous chondrites\cite{furukawa_extraterrestrial_2019} suggest that at least some sugar precursors would be available in materials similar to those comprising ocean worlds in our solar system.}

Beyond Earth, nowhere else in the solar system is the evidence for ongoing water-rock interaction stronger than at Saturn's moon Enceladus: tidal interactions support actively erupting vents at the south pole, which expel material from below the surface, directly or indirectly connected to the liquid water ocean \cite{nimmo_ocean_2016, choblet_powering_2017, kite_sustained_2016, nakajima_controlled_2016, hurford_eruptions_2007, hansen_water_2008, hemingway_cascading_2020}. Furthermore, the Cassini spacecraft provided strong evidence of water-rock interactions at Enceladus with the discovery of ammonia, methane, silica nanoparticles, hydrogen, salts, phosphates, HCN, $\ce{C2H2}$, $\ce{C3H6}$, $\ce{C2H6}$, and even macromolecular organics in the erupted plumes \cite{postberg_macromolecular_2018, Spencer06, Waite17, Postberg09, Hsu15, postberg_detection_2023, Peter24}. These latter hydrocarbons and organics could potentially contribute to the formation of complex organic molecules that could be important for the origin of life \cite{affholder_bayesian_2021}.

Similarly, other icy moons in the solar system, such as Ganymede, Titan and Europa, are likely to harbour liquid water oceans where biologically relevant organics, and particularly, components of the TCA cycle could potentially be abiotically synthesized \citep[e.g.,][]{hand_energy_2007}. The presence of methane in Titan's atmosphere \cite{raulin_organic_2002, niemann_composition_2010} strongly suggests it is being replenished from an interior source \cite{lorenz_titans_2008, sotin_observations_2012, tobie_episodic_2006}, given its rapid depletion by photolysis \cite{krasnopolsky_photochemical_2009}. On the other hand, abiotic organic synthesis in the ocean may be impeded by inefficient downward transport of organics from the ice shell into the ocean by impacts \cite{neish_organic_2024}, and a high pressure ice layer mantling the seafloor and preventing water-rock interaction, although recent models suggest that convection within the high pressure ice layer would still allow water-rock interaction to proceed \cite{kalousova_dynamics_2020, choblet_heat_2017}. Similarly, Ganymede is likely to have a high pressure ice layer mantling the seafloor \cite{journaux_large_2020}, which could convect and allow chemical exchange between the ocean and rocky interior \cite{kalousova_two-phase_2018}.

In order to investigate whether the TCA cycle is able to operate under the physical conditions relevant to solar system ocean worlds, we are informed by recent work. Canovas \& Shock \cite{Canovas+Shock20} proposed a model for the energetics of the TCA cycle at high pressure and high temperature conditions, corresponding to those of the shallow portions of Earth's subduction zones, where potential microbial habitats could exist. They explored the pressure---temperature (PT) intervals covered by the revised Helgeson-Kirkham-Flowers (HKF) model for aqueous ions and complexes \cite{tanger_calculation_1988, helgeson_theoretical_1981, johnson_supcrt92_1992, Shock92, miron_thermodynamic_2019} and the Deep Earth Water model (DEW) \cite{pan_dielectric_2013, sverjensky_water_2014, sverjensky_thermodynamic_2019, huang_extended_2019} for higher PT, and investigated the affinities of reactions in the TCA cycle from basic geochemical compounds such as \ce{CO2} and \ce{H2}. In particular, in the low T and high P parts of the deep biosphere, the overall TCA cycle released energy in the reverse direction, as long as the minimum affinities (in kcal mol$^{-1}$) were taken. With maximum affinities, a larger part of the PT domain including a subduction zone geotherm favored the forward TCA cycle as producing energy. This study highlighted the importance of exploring the energetics of biochemical reaction networks in deep habitats, since the favorability of each of the reactions forming components in the TCA cycle is dependent on the environmental conditions.

Similarly, \citep[Robinson et al.][]{Robinson21} performed experiments and modeled the formation of amide bonds with the revised HKF and DEW models for a wide range of T, P and pH, since amide bond formation leading to peptides plays a crucial role in the formation of proteins essential for life. They calculated the equilibrium constants for acetamide and diglycine formation at near-ambient pressures and 30 kbar. Intriguingly, peptide formation was found to be more efficient under high pressure suggesting that possible hydrothermal environments in ocean world interiors are favorable for the emergence of life. 

Motivated by such recent efforts, we carry out a computational study for the (non-)equilibrium {thermodynamics} of the TCA cycle and the associated prebiotic network as shown in Fig.~\ref{fig:network}, to reveal tendencies of the involved reactions  to accumulate or deplete their chemical species under deep ocean-world conditions. We investigate the stability of the reactant and product species in the constituent reactions, over a wide \emph{PT} range, covering much of the predicted range for the aqueous interiors of ocean worlds in our solar system. {In particular, we calculate the thermodynamic equilibrium states and the non-equilibrium chemical affinities of each reaction in the TCA cycle and the pyruvate-centered prebiotic network, over the \emph{PT} profiles for the interiors of Titan, Ganymede, Europa and Enceladus.}
We also investigate the {(non)equilibrium} thermodynamics of the same reactions under the \emph{PT} conditions within the Atlantis Massif, near the Lost City Hydrothermal Field in the Atlantic ocean, as a point of comparison with a location where a deep biosphere is likely to be present, utilizing species in the TCA cycle for metabolism. 

The Gibbs free energy changes and the chemical affinities of individual \emph{reactions} are often the products of computed equilibrium {and non-equilibrium} conditions, as reported in the literature (e.g. Canovas and Shock\cite{Canovas+Shock20}, Robinson et al.\cite{Robinson21}, etc.). Here, we {focus on estimating sources and sinks of individual \emph{species}, by} suggesting a new approach to quantify the stability of a given chemical species as a function of $P$ and $T$. This approach accounts for multiple reactions that involve the species, such as those encountered in a branching network with multiple sources and sinks typical of biochemical reaction networks. {To obtain a simple proxy for the average expected concentration of a given species in the network, we introduce the net chemical affinity as the algebraic sum of the affinities of the reactions connecting to that species. Positive and negative net affinities thus indicate a convergence (accumulation) and} divergence (depletion) of the species, respectively, as reactions involving that species progress.
In this way, we suggest a simple proxy indicating expected relative proportions of species involved in a given reaction network under the pressure, temperature and pH conditions of a given aqueous environment.
Based on these calculations, we 
also estimate the relative contributions of each reaction in the cycle to the total Gibbs free energy budget { (i.e., assuming independent equilibrium states)}, which may be exploited by potential organisms at these locations.

{Our analysis in this study is conducted in two distinct approaches. We first use E. coli intra-cellular concentrations under neutrality as a reference case for analyzing how these reactions might operate in a biological context. The purpose of this approach is to make an initiative to evaluate the possibility of these reactions to support the metabolism under various ocean world conditions. In the second approach, we use inorganic concentrations constrained by Cassini measurements of Enceladus, work out the equilibrium concentrations of relevant species by modelling formation reactions, also estimating organic activities from published meteoritic measurements. In this way, we assess the sensitivity of our analysis in the first approach on compositional differences, in addition to estimating Enceladus-specific net reaction affinities. 
}

\section{Methods}
\label{sec:model}

\begin{figure*}
    \centering
    \includegraphics[width=\linewidth]{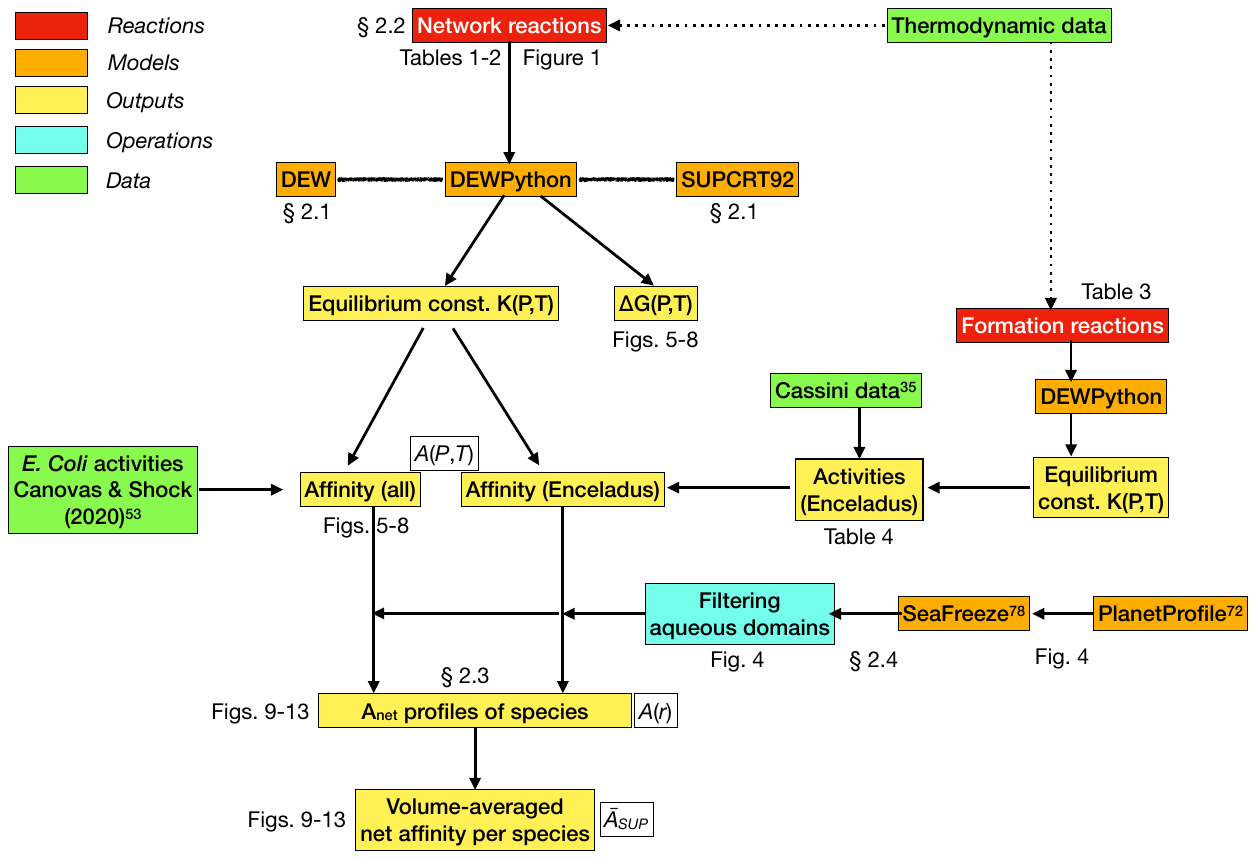}
    \caption{{Pipeline diagram for all the calculations and analysis throughout the paper. Colors denote various object classes, indicated in the legend. Relevant figures, tables, and method sections are also referenced.}}
    \label{fig:pipeline}
\end{figure*}

{Our modeling framework is depicted in Figure~\ref{fig:pipeline}. The thermodynamic models 
and the considered reactions are outlined in Sect.~\ref{ssec:thermo-models} and \ref{ssec:reac}. The net affinity as a stability metric is introduced in Sect.~\ref{ssec:species}. The procedure for determining the planetary domains for which water can be in liquid phase is presented in Sect.~\ref{ssec:waterphases}. }

\subsection{Thermodynamic models and PT ranges}
\label{ssec:thermo-models}
The Deep Earth Water model \citep[DEW][]{pan_dielectric_2013, sverjensky_water_2014, sverjensky_thermodynamic_2019, huang_extended_2019} and SUPCRT code \citep{johnson_supcrt92_1992, zimmer_supcrtbl_2016} were used in the study. The open source code DEWPython \citep{chan_dewpython_2021, chan_chandr3wdewpython_2021}, developed as a Python programing language adaptation of the 2019 DEW spreadsheet was used.\footnote{\url{https://github.com/mmelwani/DEWPython}. \textcolor{red}{Note to reviewers: this repository will be mirrored on \url{https://github.com/NASA-Planetary-Science/} after acceptance.}} 
This model calculates the molal thermodynamic properties of water, aqueous species, gases and minerals at high T (373--1473 K) and P (1--60 kbar) using the extended semi-empirical Helgeson-Kirkham-Flowers (HKF) equations of state to obtain thermodynamic quantities such as Gibbs free energy of reactions, equilibrium constant and volume change for chemical reactions that occur during water-rock interaction. SUPCRT is a code that calculates the same quantities at lower P and T (1--5000 bars and 0--1000$^\circ$C)\citep{johnson_supcrt92_1992}. In both models, we adopted thermodynamic quantities from data files provided with the DEW model and SUPCRT96. (SUPCRT96 includes improvements to heat capacity calculations over SUPCRT92, derived from Holland and Powell\cite{holland_enlarged_1990}.) For the majority of the chemical species considered, the thermodynamic quantities required for equilibrium calculations are from Canovas and Shock\cite{Canovas+Shock16}, Amend and Plyasunov\cite{Amend01}, Dick et al.\cite{Dick06} and Shock\cite{Shock95}.  

In our study, the calculations with the DEW model predominantly apply to the deep mantles of the target satellites, while the calculations with SUPCRT largely apply to the subsurface oceans.
In addition to these two models, PT profiles of Europa, Ganymede, Titan and Enceladus were computed with the PlanetProfile code \cite{styczinski_planetprofile_2023}, to be fed into SUPCRT and DEWPython to compute the thermodynamic quantities. PlanetProfile is an open source code used to perform self-consistent interior structure modeling of planetary bodies, and is specially tailored to ocean worlds and rocky planets in our solar system. For ease and consistency, the PT profiles we used are the default outputs of PlanetProfile, reported in the paper and archived on Zenodo\cite{styczinski_planetprofile_2022}. The PT profile for the Atlantis Massif near the Lost City hydrothermal field was calculated from the density, depth and temperature data logged by downhole measurements of Hole U1309D in expedition 340T of the Integrated Ocean Drilling Program (IODP) in 2012 \cite{expedition_340t_scientists_integrated_2012}. Raw and computed data for the borehole measurement are included in the \textbf{Supporting Information}.

\subsection{Model reactions}
\label{ssec:reac}

We considered the abiotic TCA cycle in the form given by \citet{Canovas+Shock16}. We have made this specific choice to make sure that all the chemical species involved were included in the thermodynamic database used by the DEW and SUPCRT models. 
In addition, we modeled an associated small `prebiotic network', at the center of which lies pyruvate. We also limited the species in the prebiotic network to those found in the thermodynamic data of aqueous species used by DEW and SUPCRT. Figure~\ref{fig:network} {shows a schematic diagram of the entire network in consideration.} 
Tables~\ref{tab:prebiotic} and \ref{tab:TCA} show the individual reactions of the prebiotic network and the TCA cycle, respectively. 


\begin{table*}
    \centering
    \begin{tabular}{ll}
    \hline\hline 
    Glucose to pyruvate& \begin{tabular}[l]{@{}l@{}} 
        $\ce{glucose} + \ce{2ADP^{3-}} + \ce{2PO4^{3-}}\ \ce{<=>}$ \\ $2~\ce{pyruvate^{-}} + \ce{2H^{+}} +  \ce{2ATP^{4-}} + \ce{2H2O}$  \end{tabular} \\
    \hline 
    Pyruvate to alanine& $\ce{pyruvate^{-}} + \ce{NADH} + \ce{H^{+}}+ \ce{NH3} \ce{<=>} \ce{alanine} + \ce{NAD^{+}} + \ce{OH^{-}}$\\
    \hline 
    Pyruvate to lactate& $\ce{pyruvate^{-}} + \ce{NADH} + \ce{H^{+}} \ce{<=>} \ce{lactate^{-}} + \ce{NAD^{+}}$\\
    \hline   
    Pyruvate to oxaloacetate& \begin{tabular}[l]{@{}l@{}} 
        $\ce{pyruvate^{-}} +  \ce{HCO3^{-}} + \ce{ATP^{4-}} \ce{<=>}$ \\ $\ce{oxaloacetate^{2-}} + \ce{ADP^{3-}} + \ce{PO4^{3-}}$ \end{tabular} \\
    \hline   
    Pyruvate to acetate&
       $\ce{pyruvate^{-}} + \ce{ADP^{3-}} + \ce{PO4^{3-}} \ce{<=>} \ce{acetate^{-}} + \ce{ATP^{4-}} + \ce{CO2}$\\
    \hline   
    Acetate to citrate& \begin{tabular}[l]{@{}l@{}} 
        $\ce{acetate^{-}} +  \ce{oxaloacetate^{2-}} + \ce{ATP^{4-}} + \ce{H2O} \ce{<=>}$ \\ $\ce{citrate^{3-}} + \ce{ADP^{3-}} + \ce{PO4^{3-}}$ \end{tabular} \\
    \hline 
    \end{tabular}
    \caption{Prebiotic network reactions modeled, after Nelson and Cox\cite{Nelson2017}.}
    \label{tab:prebiotic}
\end{table*}


\begin{table*}
    \centering
    \begin{tabular}{ll}
    \hline\hline 
    Step 0&\begin{tabular}[l]{@{}l@{}}
        ${\rm pyruvate^{-}} + \ce{NAD^{-}ox}+{\rm oxaloacetate^{2-}}+ \ce{H2O} \ce{<=>}$\\ ${citrate^{3-}}+ \ce{NAD^{2-}red} + \ce{CO2} + \ce{H^{+}}$ \end{tabular}\\\hline 
    Step 1& $\ce{citrate^{3-}} \ce{<=>} cis{\rm -aconitate^{3-}} + \ce{H2O}$\\
    \hline 
    Step 2& $cis{\rm -aconitate^{3-}} + \ce{H2O} \ce{<=>} \ce{isocitrate^{3-}}$\\
    \hline 
    Step 3&  
        \begin{tabular}[c]{@{}l@{}}
        ${\rm isocitrate^{3-}} + \ce{NAD^{-}ox} \ce{<=>}$ \\ $\alpha{\rm -ketoglutarate^{2-}} + \ce{NAD^{2-}red} + \ce{CO2}$ 
        \end{tabular}\\
    \hline 
    Step 4&
        \begin{tabular}[l]{@{}l@{}}
        $\alpha{\rm -ketoglutarate^{2-}} + \ce{NAD^{-}ox} + \ce{ADP^{3-}} + \ce{HPO4^{2-}} \ce{<=>}$ \\
        $\ce{succinate^{2-}}  + \ce{NAD^{2-}red} + \ce{CO2} + \ce{ATP^{4-}}$
        \end{tabular}\\
    \hline
    Step 5&
        \begin{tabular}[l]{@{}l@{}}
        ${\rm succinate^{2-}} \ce{<=>}{\rm fumarate^{2-}} +\ce{H2}$ \end{tabular}\\ \hline 
 Step 6&\begin{tabular}[c]{@{}l@{}}
        ${\rm fumarate^{2-}}+ \ce{H2O}  \ce{<=>} {\rm malate^{2-}} $\end{tabular}\\ \hline 
 Step 7&\begin{tabular}[c]{@{}l@{}}
        ${\rm malate^{2-}} + \ce{NAD^{-}ox} \ce{<=>} {\rm oxaloacetate^{2-}} + \ce{NAD^{2-}red} + \ce{H^{+}}$ 
        \end{tabular}\\ \hline
    \end{tabular}
    \caption{Steps of the TCA cycle modeled, after Canovas and Shock\cite{Canovas+Shock16}.}
    \label{tab:TCA}
\end{table*}

{We approach the reaction network in Table~\ref{tab:prebiotic} with careful consideration of recent advances in prebiotic chemistry. While biological cofactors like ATP and NADH are traditionally associated with enzyme-mediated metabolism, mounting evidence suggests their functionality and relevance in prebiotic contexts. Recent experimental work has demonstrated that NAD(H) can be reactive in protocellular environments with photochemical reactions with and without FeS clusters and other reactants\cite{Summers2015,Bonfio2018,Dalai2020}. In addition, NAD has been demonstrated to be reactive with simple FeS mineral species, $\alpha$-keto acids and \ce{H2} gas\cite{Basak2021,Weber2022,HenriquesPereira22}. In addition, a prebiotic synthesis for NAD has been proposed by Kim and Benner.\cite{Kim2018} ATP has been less studied within the prebiotic context; however, computational work by Chu \& Zhang\cite{Chu2023} suggests that volcanic phosphate minerals could thermodynamically drive ATP synthesis in hydrothermal environments. These studies collectively support the plausibility of ATP and NADH participating in protometabolic networks before the evolution of sophisticated enzymatic systems. In our modeling, we consider these cofactors primarily for their thermodynamic contributions to reaction stoichiometry rather than their evolved biological functions.}

\subsection{Assessing species stability}
\label{ssec:stab}

\begin{figure}
    \centering
    \includegraphics[width=\linewidth]{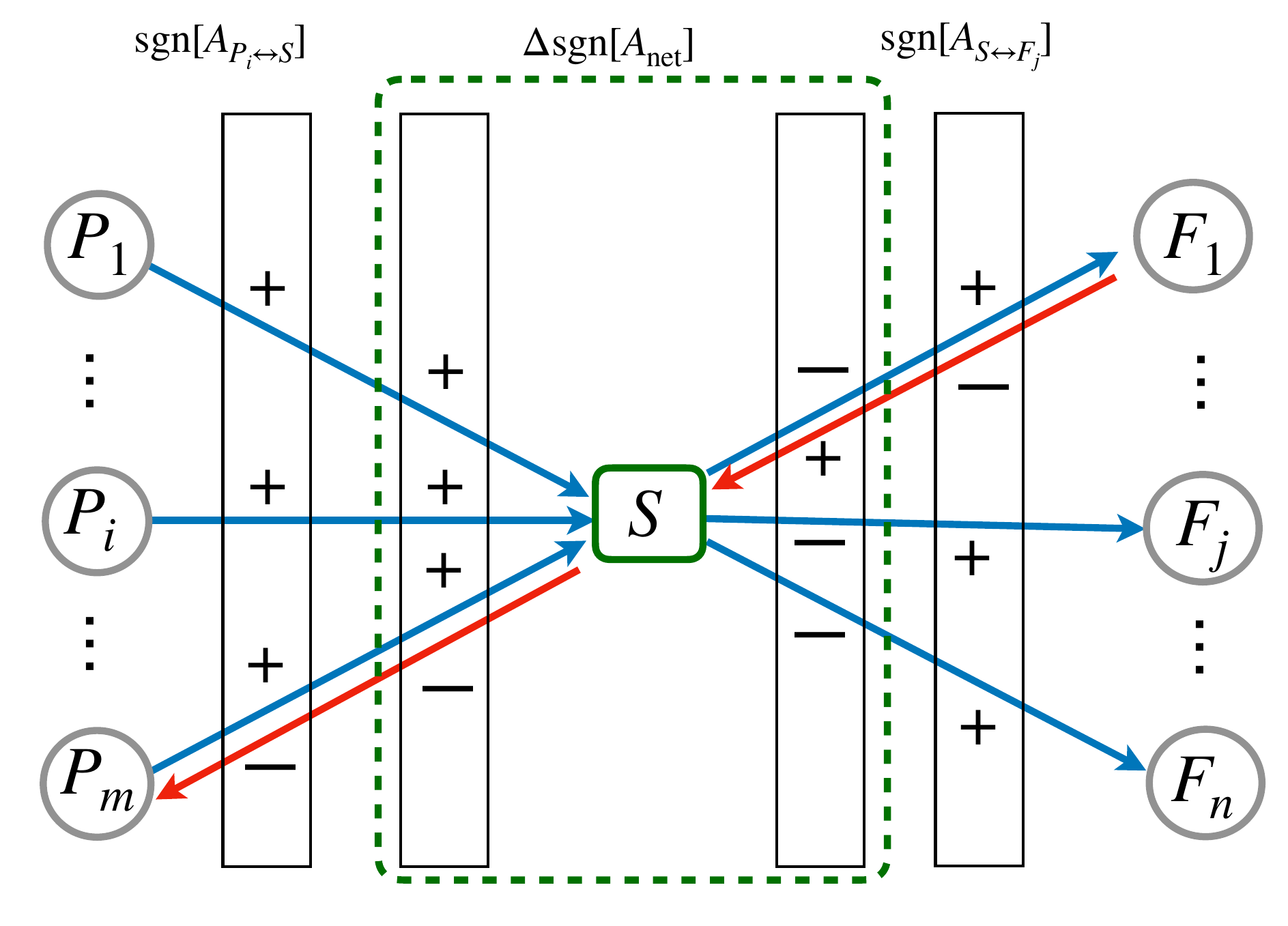}
    \caption{A graphical representation of a section in the reaction network, centered on species $S$. The node species, $S$, is either the reactant or the product of the edge reactions indicated with the conventional direction. The edges are blue (red) when they are in the conventional (anti-conventional) direction. For a given pressure and temperature, {the net affinity} is either {convergent} (positive) or {divergent} (negative), as indicated in the figure. The {contributions to the} sign (``sgn'') of the net {affinity} is depicted around the central species, $S$.}
    \label{fig:sketch}
\end{figure}

{In this section,} we {set up a quantitative measure of} the expected {energetic feasibility for the net production of a given} chemical species as a function of 
P and T, {based on the chemical affinities of those reactions which involve the species.} 
We consider the {reaction} network 
as a graph, where the nodes and edges represent the species and reactions, 
respectively. A section of such a network is shown in Fig.~\ref{fig:sketch}, centered on a species $S$. 

To quantify the thermodynamic driving forces acting on chemical species in ocean world environments, we assess both equilibrium conditions and non-equilibrium chemical affinities. The affinity $A$ measures the energy that would be instantaneously released when a reaction proceeds from its current state (i.e., with certain concentrations of reactants and products) toward equilibrium. It is defined as the negative derivative of the Gibbs free energy with respect to reaction progress, $A := -\partial G / \partial\xi$. Since we calculate equilibrium constants $K$ from standard Gibbs energy changes ($\Delta G$) using: 
\begin{equation}
    log_{10}K = -\frac{\Delta G}{2.303 RT}, 
\end{equation}
we can determine the affinity of reactions starting from any distribution of concentrations through:
\begin{equation}
    A = 2.3026 RT(\log_{10} K - \log_{10} Q).
    \label{eq:affinity}
\end{equation}
The affinity differs from $\Delta G$ by the additive term $2.3026RT\log Q$, where $Q$ is the reaction quotient, which depends on the activities $a_i$ and the stoichiometric coefficients $s_i$ of the constituent products $i$ and reactants $j$:
\begin{equation}
    \log Q = \sum_{i}(s_i\log a_i) - \sum_{j}(s_j\log a_j). 
\end{equation}
{We follow two distinct approaches when assuming the discrete distribution of activities for the chemical species considered. Firstly (Sect.~\ref{ssec:Ecoli}),} taking a similar approach to Canovas \& Shock\citep{Canovas+Shock20}, we use intra-cellular data from E. Coli bacteria and extra-cellular data as well(\citep[][see their Table 25.1]{Bennett2009, Canovas+Shock20}). 
{We give the E. Coli activities of all the species involved in the network reactions in the Supporting Information.}

{In this way, we carry out a type of ``thermal stress test" for the reactions in the network: by taking concentrations we know work in a biological system and subjecting them to ocean world conditions, we explore how robust these reaction pathways might be under different pressure-temperature regimes. This approach helps us understand whether these reactions could potentially serve as a foundation for metabolism under ocean world conditions, even if the actual concentrations would differ in an abiotic setting.
In addition, we also use measurements and estimates of species activities specific to Enceladus to calculate affinities (Sect.~\ref{ssec:enceladus}), providing a direct comparison between our biologically-based reference case and a more realistic abiotic scenario based mainly on meteoritic and Enceladus-specific compositions.}

{{To quantify the effect of multiple reactions on the instantaneous change in a given species' concentration,} we aggregate information from the edge reactions connected to a given node species $S$, by taking reaction affinities into an algebraic sum. This is shown schematically in Fig.~\ref{fig:sketch}.
}
Here, the classes of species $P$ and $F$ are, respectively, the presumed 
sources and sinks to the species of interest, $S$, which we call the preceding 
($P$) and the following ($F$) species. 
We choose the conventional direction\footnote{We use the term conventional here 
to indicate the biologically expected forward direction.} for each reaction in the 
network to be the expected dominant direction, which puts the main reactant or product 
to either the preceding or the following class. 
We thus quantify the {net affinity toward} $S$ as the {difference} 
between the {algebraic sums} of {affinities aggregated from} the series $P$ and the series 
$F$, so that 
\begin{eqnarray}
    A^{({\rm S})}_{\rm net} &=& \sum_i A_{\rm P_i\leftrightarrow S} - \sum_j A_{\rm S\leftrightarrow F_j} \nonumber \\
    &=& \sum_k \nu_k A_k
    \label{eq:Anet}
\end{eqnarray}
where ${\rm P}_i$ and ${\rm F}_j$ are the $i$th and $j$th species on the 
corresponding classes.
{For a species $S$ appearing with a sign factor $\nu_k$ (+1 or -1) in reaction $k$, the contribution to $A_{\rm net}$ is weighted as $\nu_k A_k$. The sign of $\nu_k$ distinguishes whether $S$ appears as reactant (negative) or product (positive).}
{If $A^{({\rm S})}_{\rm net}$ is positive (negative), then more (less) energy would be produced in reactions that flow toward $S$, than away from $S$. This indicates that, when no energy is provided externally, the involved reactions will spontaneously increase ($A_{\rm net}>0$) or decrease ($A_{\rm net}<0$) the concentration of species $S$. In this way, the net affinity toward $S$ can be seen as a proxy for the expected abundance of $S$, provided that the assumed concentrations of $P_i$, $F_j$, and the other constituents of the edge reactions connecting to $S$ are known \emph{a priori}. We note that the net affinity is an `instantaneous' quantity, in the sense that it provides with a snapshot abundance proxy. $A_{\rm net}$ combines the affinities of reactions, which in general host more than just the species in question. This summation of affinities assumes that the initial rates of individual reactions are independent, that changes in species concentrations are small enough that affinities remain near their initial values, and that cross-coupling between reaction rates can be neglected in the initial instant. While $A_{\rm net}$ provides insight into the initial thermodynamic driving forces, the actual temporal evolution of species concentrations would require solving the coupled reaction kinetics equations, which is beyond the scope of this study.}

\subsection{Water phases along PT profiles, and considering reactions in liquid regions only}
\label{ssec:waterphases}

SUPCRT and DEW are only valid at the PT regions where liquid water is stable. However, various ice phases are predicted to be stable along the present-day geotherms of ocean worlds \citep[e.g.][]{sotin_internal_2004, vance_geophysical_2018, journaux_large_2020}. 
To determine the water phases along the PT curves to exclude icy regions from our calculations for aqueous species, we used the SeaFreeze\footnote{https://github.com/Bjournaux/SeaFreeze} code \citep{journaux_holistic_2020}, which solves the equations of state of water between 130--500 K up to 2300 MPa (23 kbar). While the PlanetProfile models also compute the thermodynamically stable water phases along the radial profiles of the bodies, and are informed by SeaFreeze for pure water oceans, the default models we use assume ocean compositions and salt concentrations which are in reality not well constrained. In addition, while PlanetProfile can include fluid-filled porosity in rocky interiors (Enceladus' structure includes $\sim$ 32 \% ocean-filled porosity at the seafloor to $\sim$ 22 \% at the center), assumptions about pore closure pressures are required, to best-fit the bodies' moments of inertia, so real porosities within these bodies are not well constrained. We therefore adopted SeaFreeze results for pure water phases overlain on the default PlanetProfile profiles in our models, to see where pure liquid water would be thermodynamically stable.

\begin{figure}
    \centering
    \includegraphics[width=1.1\linewidth]{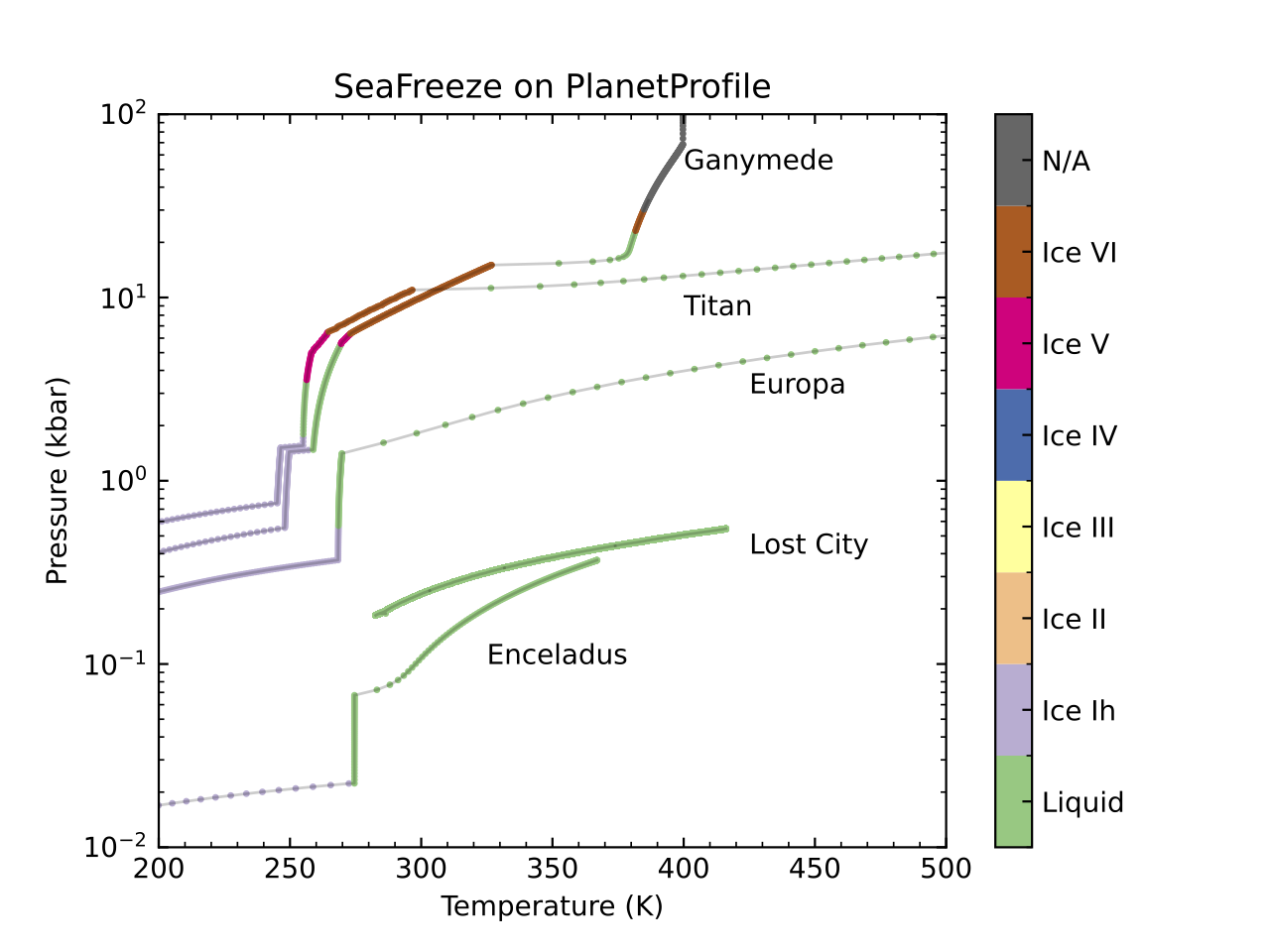} \\
    \includegraphics[width=1.1\linewidth]{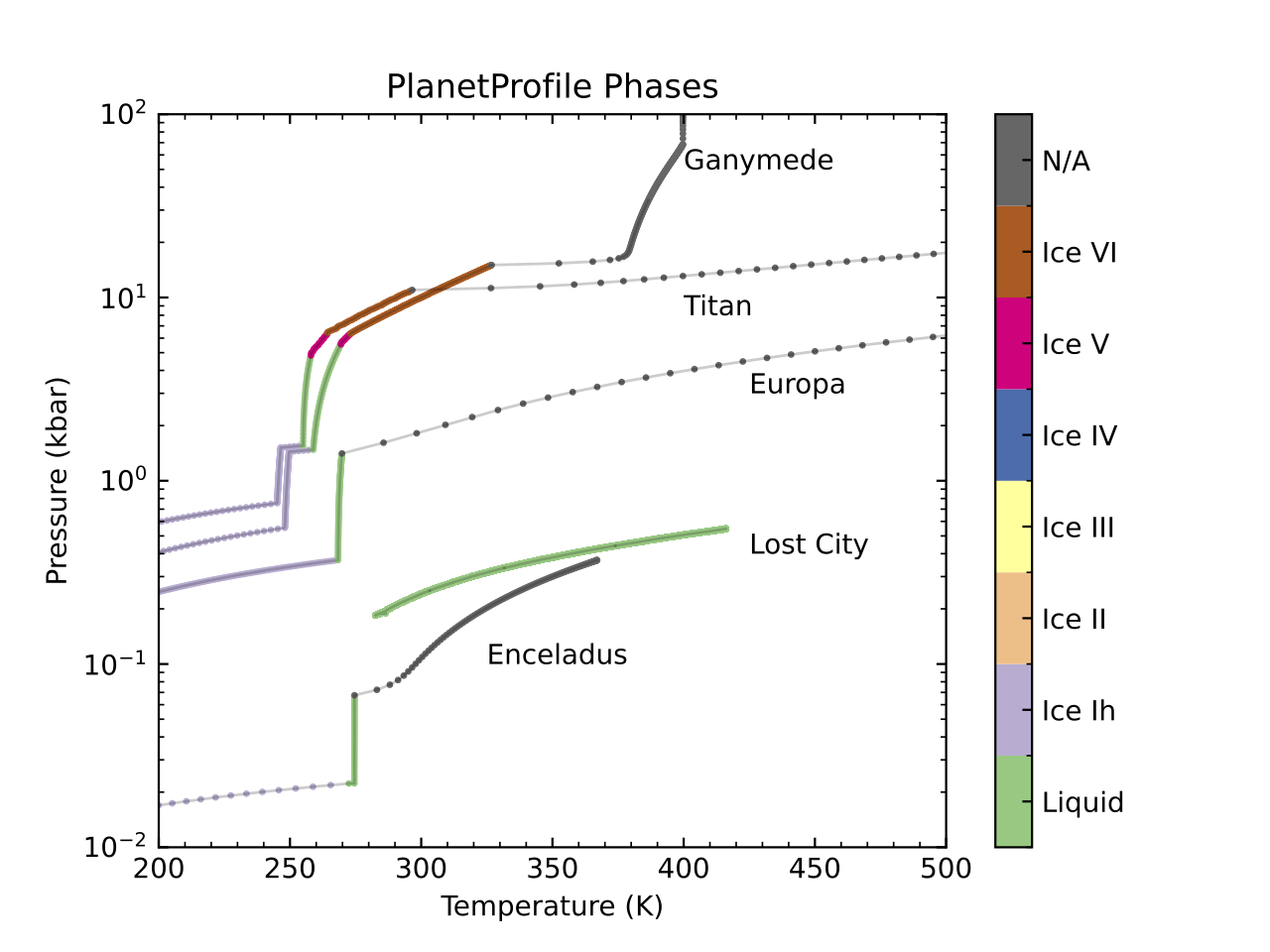}
    \caption{The predicted water phases of Enceladus, Europa, Titan, and Ganymede, based on the SeaFreeze model overlain on PlanetProfile results (top panel) and water phases according to the default PlanetProfile models, which include non-pure water oceans (bottom panel). The purple regions indicate icy outer shells, while the green regions represent areas where liquid water could exist, including oceans and pores in the rocky interior. Colors other than green signify various states of ice. All water phases included in SeaFreeze are illustrated in the color bar, including those not calculated to be stable in these bodies. N/A in the top panel refers to regions out of the range of SeaFreeze, and in the bottom panel refers to the rocky interior.}
    \label{fig:seafreeze}
\end{figure}

The resulting water phase profiles are shown in Fig.~\ref{fig:seafreeze}. The profile of Europa has a thicker icy crust than the default (salty) PlanetProfile model. The rocky layer beneath Europa's seafloor appears to be an environment where liquid water can be thermodynamically stable to significant depths (Fig.~\ref{fig:seafreeze}), provided the interior contains porosity, although the presence of porosity is under debate in ocean world interiors at high P\cite{byrne_limited_2018, klimczak_strong_2019}. If water-filled pores are present, aqueous solutes resulting from water---rock interaction would also undoubtedly be present.

In Titan and Ganymede, the ocean rests at the top of high pressure ice layers (V and VI). At Ganymede, a high pressure ice layer separates the ocean in two, and a high pressure ice layer mantles the rocky interior, which may allow liquid water if fractures or pores are present. Liquid water is stable throughout the entire interior of Enceladus, including the rocky portion. 
As a reference case from Earth's crust, we also show the PT profile based on measurements from the borehole near the Lost City Hydrothermal Field.

On the bottom panel of Fig.~\ref{fig:seafreeze}, we show for comparison the default PlanetProfile results for the water phases along the same profiles, which generally agree with the phases predicted by SeaFreeze.  

To determine the net equilibrium constant of a given species (Eq.~\ref{eq:Anet}) along the parts of the planetary body PT profiles where pure liquid water is stable, a two-dimensional linear spline interpolation was performed on the calculated $\log K_{
rm net}$ of each species on the PT plane. Extrapolated values were used along those parts of the PT curves outside the domains of DEW and SUPCRT. 
This was the case for Titan, Ganymede, and the Europan ocean. The extents of the extrapolations can be seen by comparing Fig.~\ref{fig:seafreeze} and Fig.~\ref{fig:res_prebiotic}. 
For SUPCRT, we used a rectangular domain, bounded by the minimum temperature, 273 K. The extrapolated regions are the oceans of Europa and Titan, the outer ocean of Ganymede, and the shallowest 5 km of Titan's mantle that allow for the thermodynamic stability of liquid water. The oceans have nearly constant temperature and they are stratified mainly as a function of pressure (see Fig.~\ref{fig:seafreeze}). In general, the ocean extrapolations are within a few tens of kelvins away from SUPCRT in temperature. Given that the $\log_{10}~K_{\rm net}$ contours are also nearly vertical in those regimes \citep[see also][]{Canovas+Shock20}, extrapolations can be seen as reasonable estimates. To test this, we forced the low-temperature limit of the SUPCRT model to work between 263 to 268 K for the allowed pressure range (1--3 kbar). The resulting contours showed a good match with extrapolations. We show in {Supporting Information} an example SUPCRT calculation that was forced to sub-273 K temperatures, so long as the calculation is within the applicable bounds of the model, leading to a triangular region there. We note that the extrapolated values of $K$ are in line with the forced SUPCRT model. 

\section{Results}
\label{sec:res}

{To obtain chemical affinities of the reactions in 
Tables~\ref{tab:prebiotic}-\ref{tab:TCA}, we need to specify the chemical activities of all the species involved. We organize the analysis in two distinct approaches, as introduced in Sect.~\ref{ssec:stab}. In the first one (Sect.~\ref{ssec:Ecoli}), we pose the following question. If life were to exist in these environments, could this specific reaction network and its node species be thermodynamically favorable? To provide an insight, we follow a similar approach with Canovas \& Shock\cite{Canovas+Shock20} (who focused on deep-Earth habitats) and adopt E. Coli intracellular concentrations and neutral pH as instantaneous activities for chemical affinity through ocean-world interiors, leading to net affinities per species. In the second approach (Sect.~\ref{ssec:enceladus}), we employ measured concentrations of \ce{CO2}, \ce{H2}, \ce{HPO4}, \ce{NH3}, calculate the equilibrium concentrations of some species, and convert meteoritic concentrations of the node organics of the network to solute concentrations. We then take the resulting values as activities in calculating chemical affinity, subject to Enceladean ocean and the porous rocky interior conditions and a pH of 9 to 11.} 

\subsection{{Reference analysis using E. Coli activities}}
\label{ssec:Ecoli}
In Sects.~\ref{ssec:prebiotic} and \ref{ssec:citric}, we present non-equilibrium affinities and the (equilibrium) Gibbs free energy changes for the prebiotic network reactions and the TCA cycle, respectively, covering the PT ranges chosen for DEW and SUPCRT models (Sect.~\ref{ssec:thermo-models}). In Sect.~\ref{ssec:species} we present the net affinity distributions for those species which have more than one edge in Fig.~\ref{fig:network}.

\subsubsection{Prebiotic network}
\label{ssec:prebiotic}

Figures~\ref{fig:res_prebiotic}-\ref{fig:res_prebiotic_1} show the affinity and the equilibrium change in Gibbs free energy calculated with the thermodynamic models for the prebiotic network connected to the TCA cycle. The reactions with $\Delta G < 0$ are exergonic {if they reach the equilibrium state and for $\Delta G > 0$ they are endergonic. To determine the direction of a given reaction in a non-equilibrium state, we calculated the chemical affinity (Eq.~\ref{eq:affinity}), by assuming the initial activity distribution adopted by Canovas \& Shock (2020) for E. Coli intracellular data\citep{Bennett2009}. Positive $\Delta G$ means that the reaction must be provided energy from external sources to drive it in the forward direction, to establish chemical equilibrium. However, the actual direction of the spontaneous reaction can be different with respect to what $\Delta G$
indicates, when the reaction starts from a non-equilibrium state, also depending on the ambient pH value.} 

Significantly, glycolysis is an effective pathway for the production of pyruvate for all pressures, for $T\lesssim 350$~K, i.e., throughout all the ocean PT curves. In addition, the conversion of pyruvate to alanine (predominantly in the shallow depths) and lactate (across all depths) is comparatively limited. This process would otherwise effectively divert the precursor prebiotic molecules from the TCA cycle, as neither alanine nor lactate are contributors to the cycle. Pyruvate, which typically contributes to the TCA cycle in two ways (Fig.~\ref{fig:network}), is additionally produced from oxaloacetate on the one hand and {strongly} converted to acetate on the other, which is then converted to citrate in the TCA cycle. Interestingly, the production of citrate from acetate is favored in the oceans of the ocean worlds {(see also Sect.~\ref{ssec:enceladus})}, but not at Lost City, {nor in the possibly aqueous depths of the Europan and Enceladean mantles. In the Enceladean ocean, acetate-citrate reaction is very close to zero affinity, while the equilibrium $\Delta G$ favors citrate over acetate (see also Sect.~\ref{ssec:enceladus}).} As can be seen, this prebiotic network helps the TCA cycle to by-pass the oxaloacetate node, which itself is a bottleneck of the TCA cycle through much of the potential habitats (PT curves) considered in this study (see Sects.~\ref{ssec:citric}, {\ref{ssec:species}, and \ref{ssec:enceladus}}). 

\begin{figure*}
    \centering
    \includegraphics[width=\linewidth]{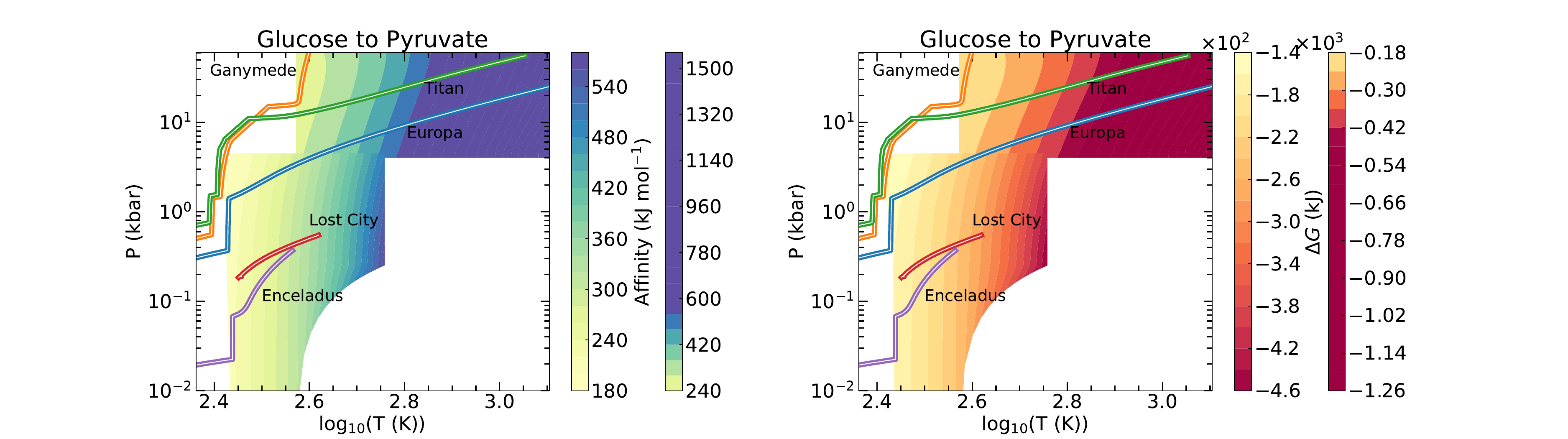}
    \includegraphics[width=\linewidth]{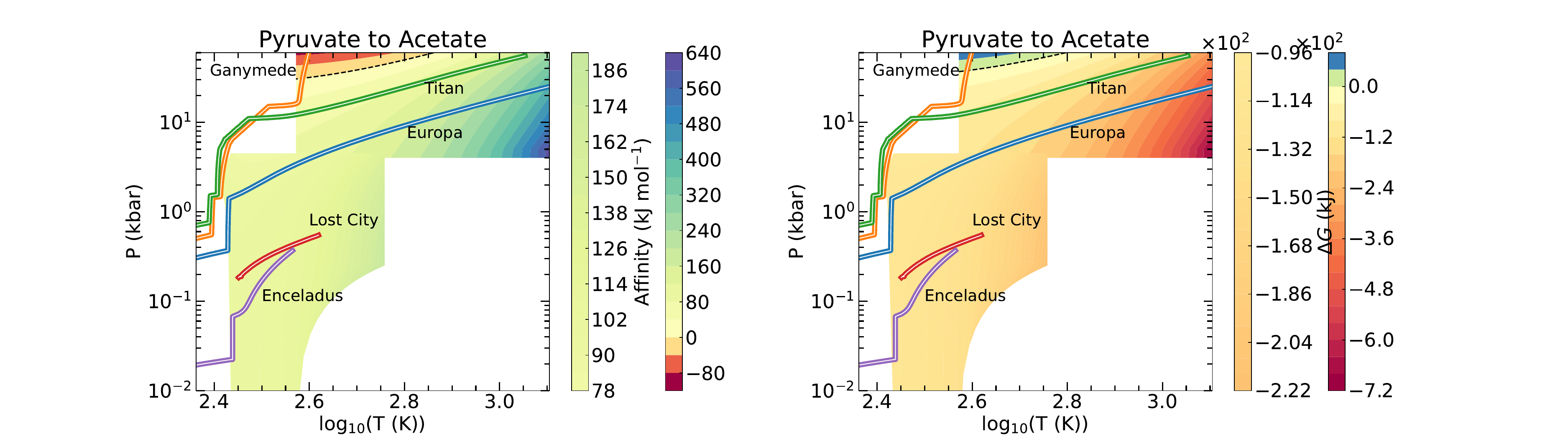}
    \includegraphics[width=\linewidth]{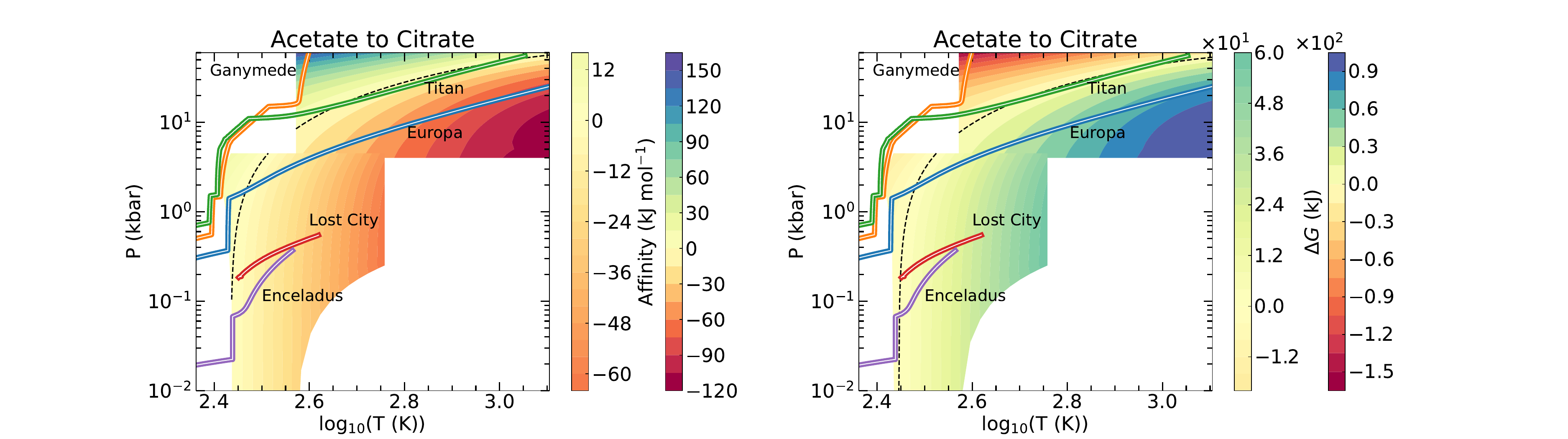}
    \caption{{Affinity} (left panels) and Gibbs free energy change (right panels) of the reactions involved in the prebiotic network. The curves show the PT curves of the studied moons and the depths near the Lost City Hydrothermal Field. The dashed curve indicates the critical {affinity} (A=0; left) or the transition between {forced-endergonic and spontaneous-exergonic reaction states in equilibrium} (right).}
    \label{fig:res_prebiotic}  
\end{figure*}
\begin{figure*}
    \centering
      \includegraphics[width=\linewidth]{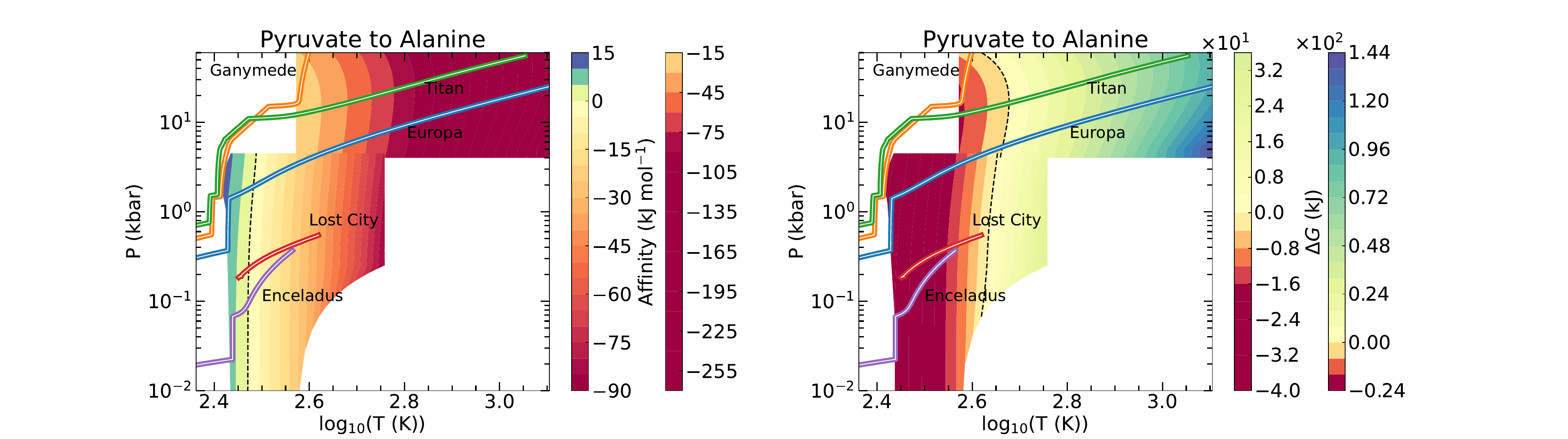}
    \includegraphics[width=\linewidth]{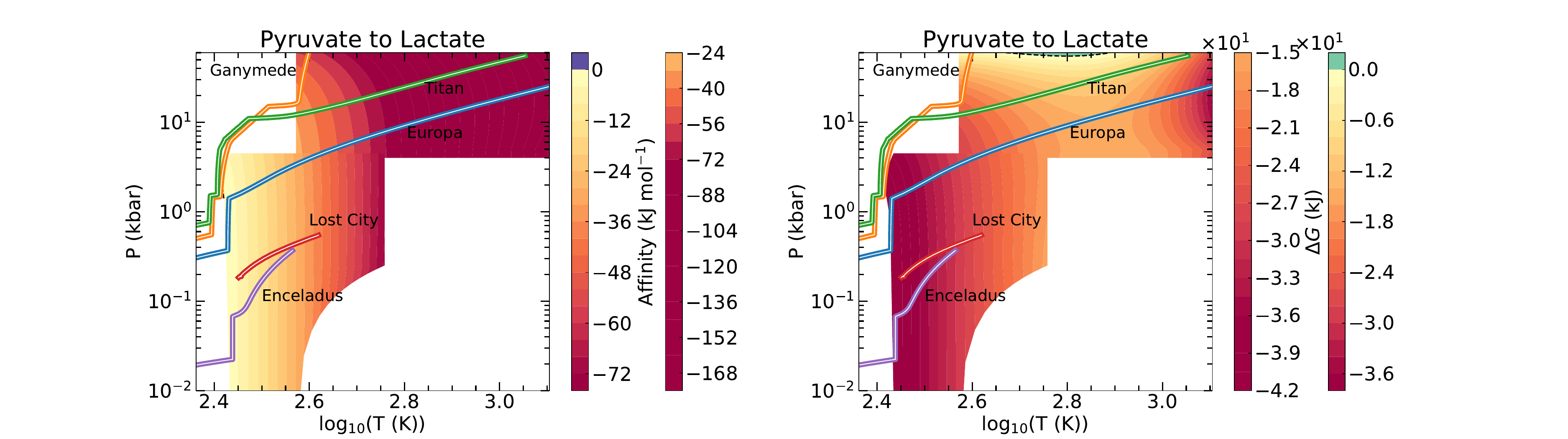}
    \includegraphics[width=\linewidth]{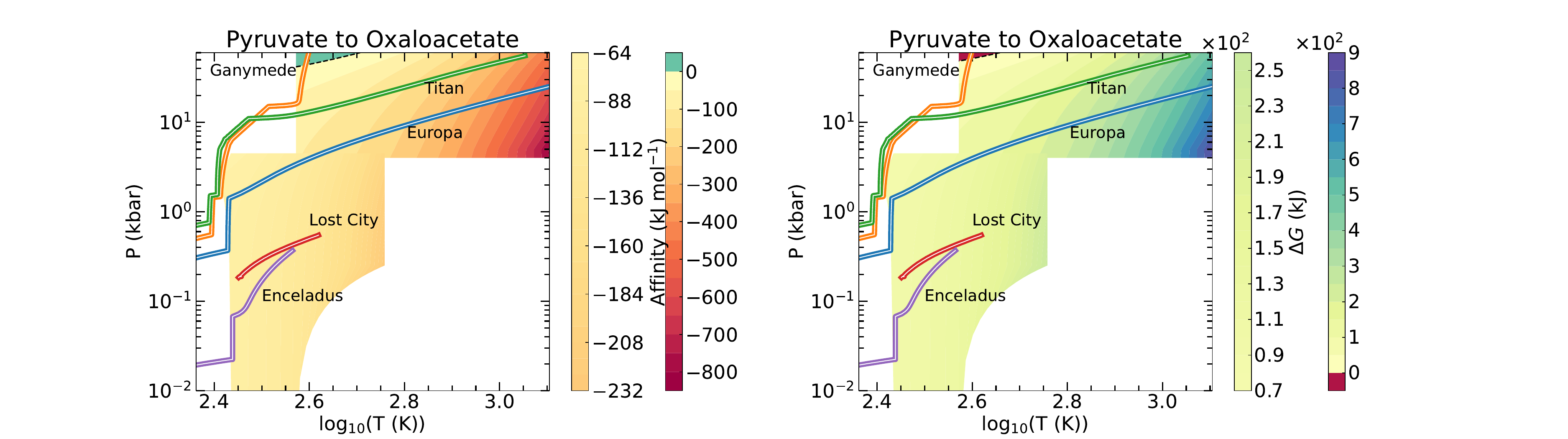}
    \caption{Continued from Fig.~\ref{fig:res_prebiotic}.}
    \label{fig:res_prebiotic_1}
\end{figure*}

\subsubsection{Citric acid cycle}
\label{ssec:citric}

Figure~\ref{fig:TCA} shows the {affinity} and Gibbs free energy change, $\Delta G$, for each reaction in the TCA cycle. 
The {Gibbs free energy changes} throughout substantial portions of all the considered PT profiles reveal a quasi-periodic directional pattern for the TCA cycle, which is otherwise expected to proceed clockwise (i.e., forward; Fig.~\ref{fig:network}) as it occurs in aerobic life as we know it. The succession of TCA cycle reactions alternates direction, proceeding forward, then reversing, and continuing in this alternating pattern. Under abiotic conditions (i.e., in the absence of enzymes), we thus see energy barriers {in states of reaction equilibria,} inhibiting the production of oxaloacetate, cis-aconitate, and fumarate throughout most of the PT profiles where liquid water would be thermodynamically stable. Production of $\alpha$-ketoglutarate appears to increase under increased T, so steeper thermal profiles (Lost City and the interiors of Enceladus and Europa) favor its production, whereas the oceans, and the interiors of Ganymede and Titan inhibit the conversion. {For} $\Delta G>0$, the inhibited reactions can be forced back to the forward direction only by the external provision of energy, e.g., provided by metabolic processes. Energy barriers exist in all the bodies of interest, including at the Lost City Hydrothermal Field. However, the bottleneck nodes seem to be different for different temperatures and pressures. For a more rigourous evaluation of relative abundances of individual node species, we need to calculate species stability under the combined actions of the edge reactions (see Sect.~\ref{ssec:stab}), which we consider in Sect.~\ref{ssec:species}. 

\begin{figure*}
    \centering
    \includegraphics[width=.9\linewidth]{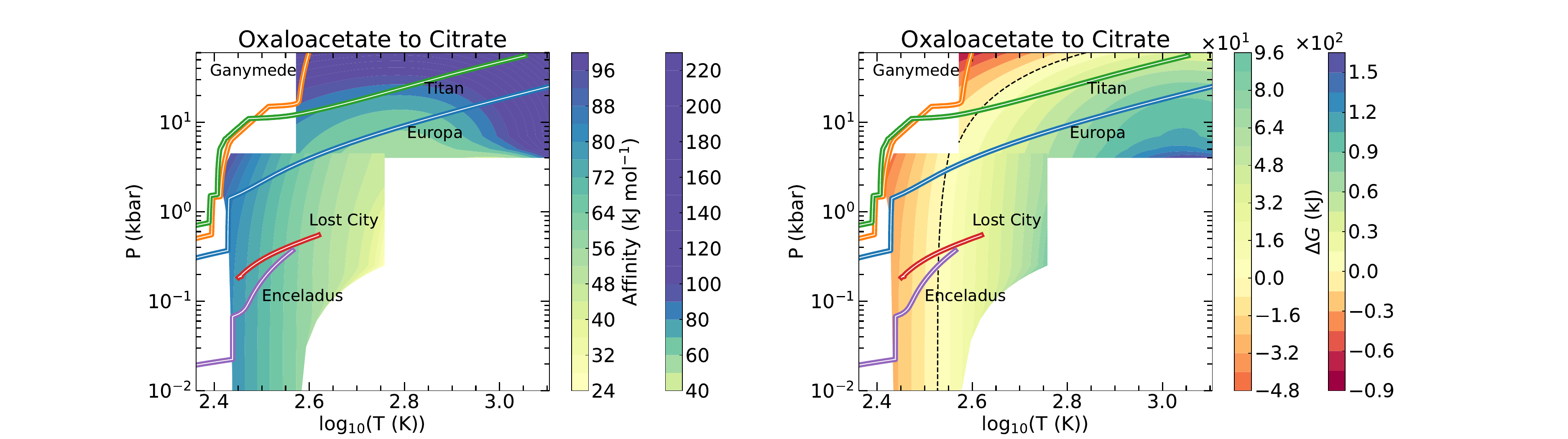}
    \includegraphics[width=.9\linewidth]{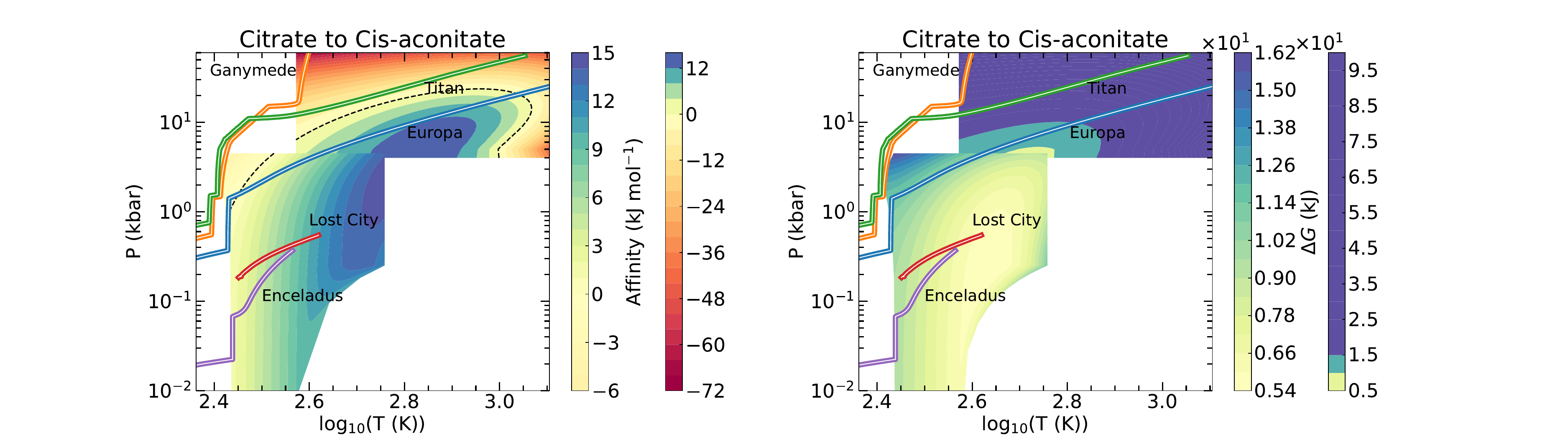}
     \includegraphics[width=.9\linewidth]{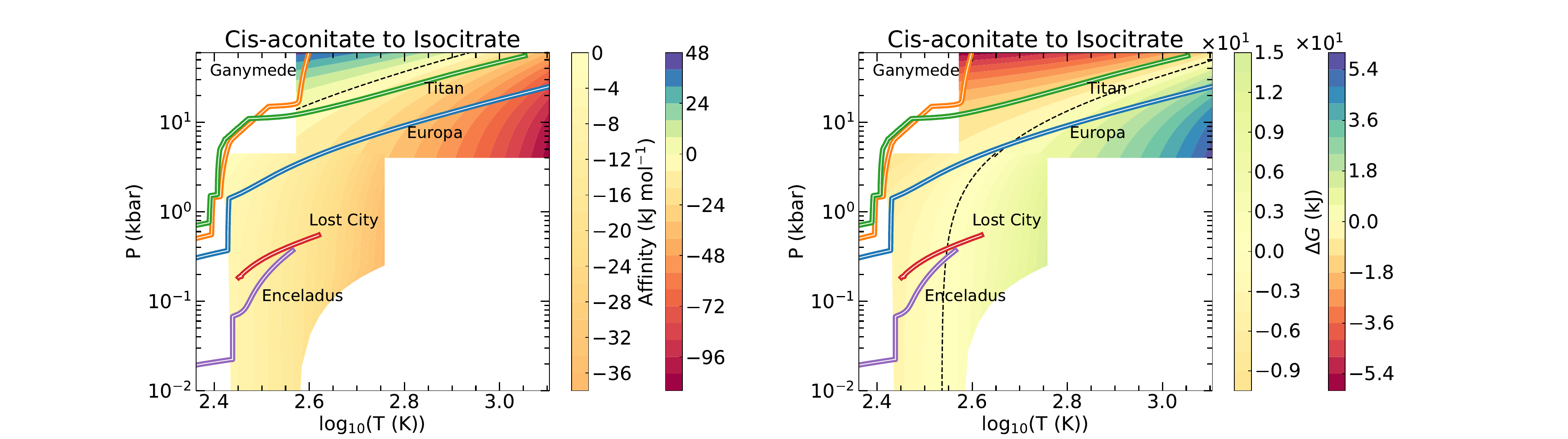}
     \includegraphics[width=.9\linewidth]{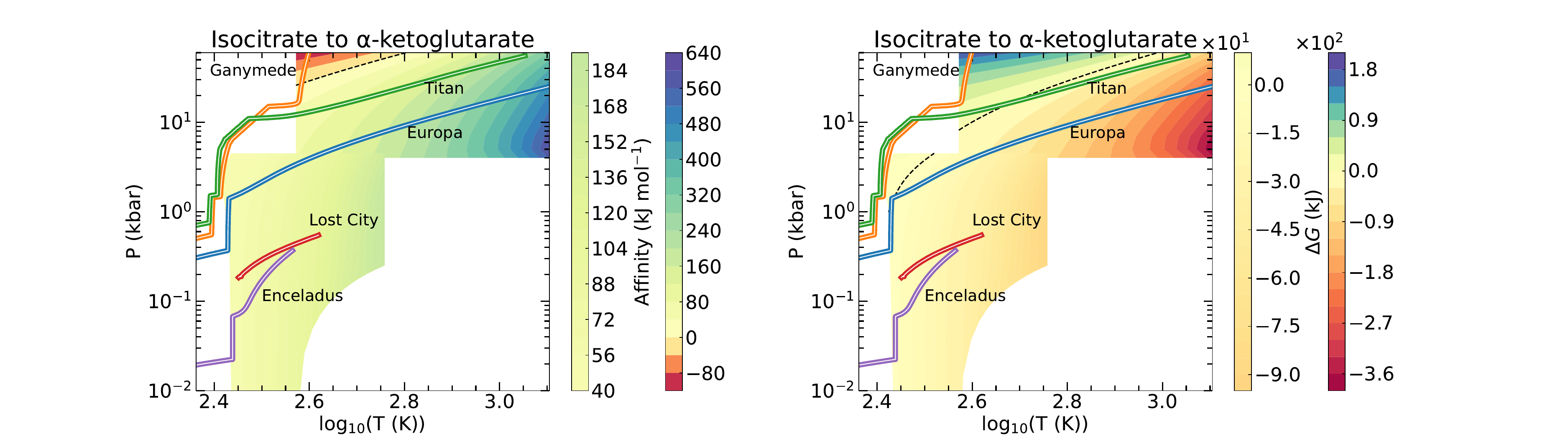}
    \caption{Stability and energetics of the forward TCA cycle on 
    the PT plane. Each horizontal plot couple shows the {affinity} (left plots) and the Gibbs free energy change (right plots) for the individual steps of the TCA cycle.}
    \label{fig:TCA}
\end{figure*}

\begin{figure*}
     \includegraphics[width=.9\linewidth]{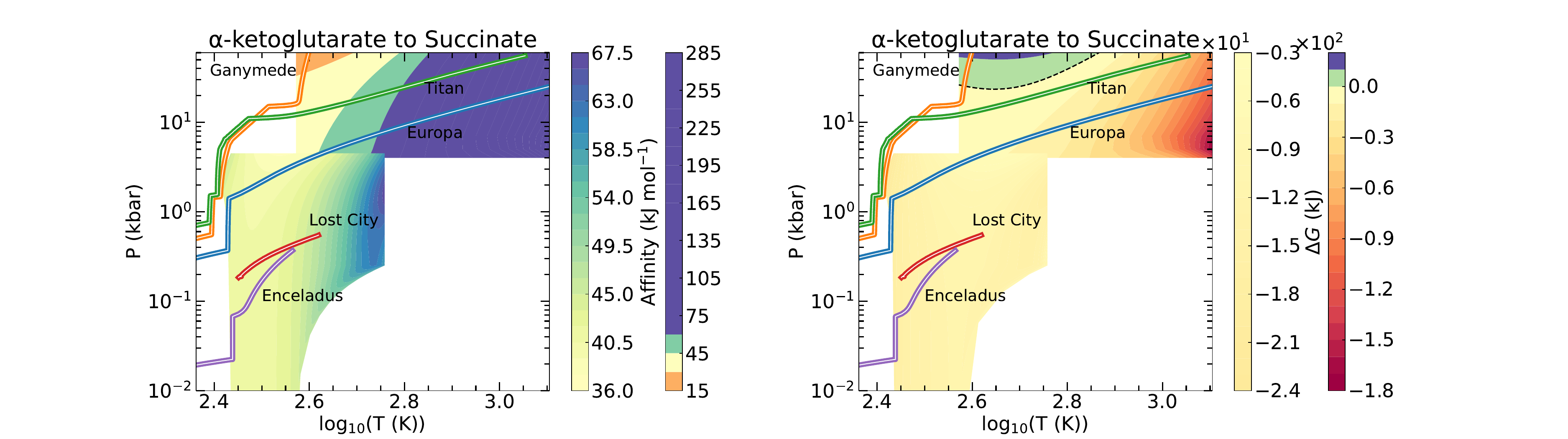}
    \includegraphics[width=.9\linewidth]{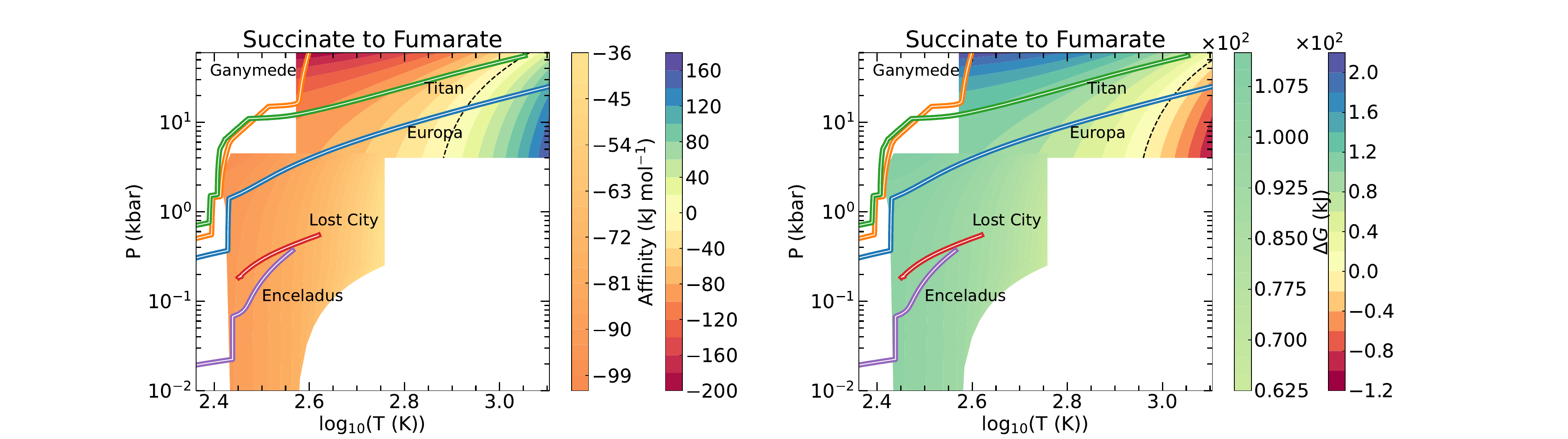} 
    \includegraphics[width=.9\linewidth]{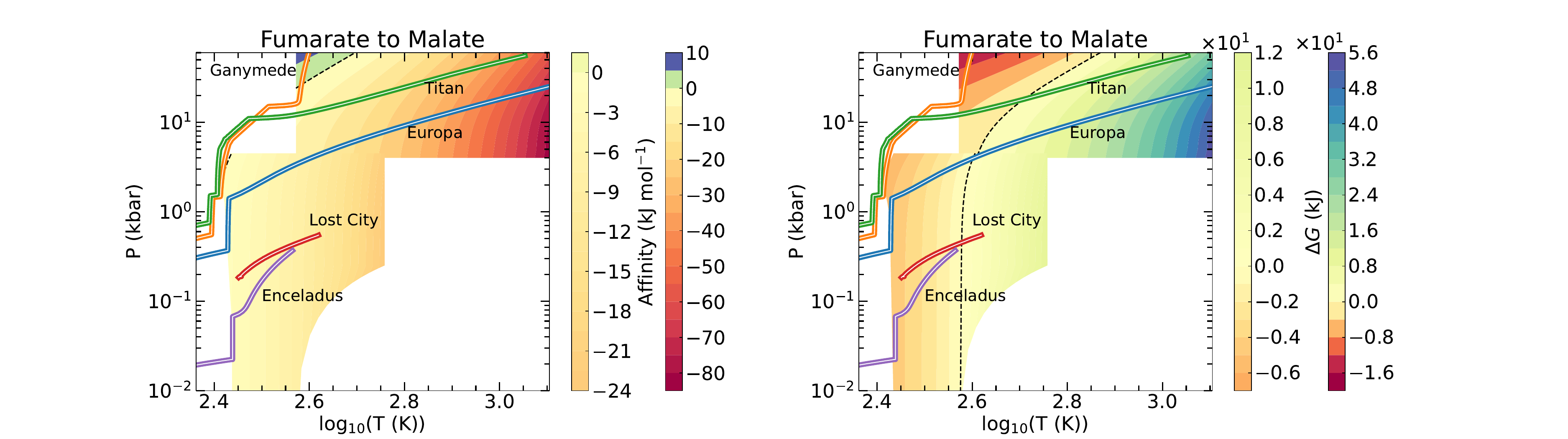}
    \includegraphics[width=.9\linewidth]{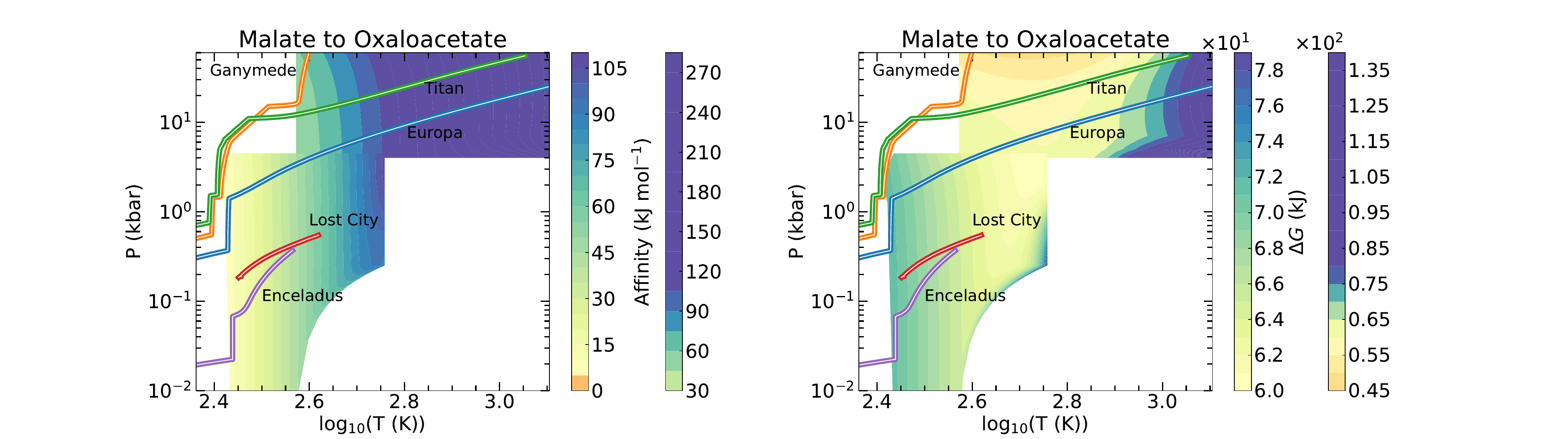}
    \caption{Continued from Fig.~\ref{fig:TCA}}
    \label{fig:TCA_1}
\end{figure*}

{This equilibrium picture of spontaneous reaction feasibility changes significantly, when non-equilibrium concentrations are considered and affinities are calculated (left panels of Figs.~\ref{fig:TCA}-\ref{fig:TCA_1}). When the non-equilibrium concentrations based on terrestrial (\emph{E. Coli}) conditions are adopted, oxaloacetate can be sourced from malate, but the former is still strongly disfavored in the network, because the other reactions consume oxaloacetate much more strongly.}

\subsubsection{Species stability}
\label{ssec:species}

\begin{figure*}
    \centering
    \includegraphics[width=0.85\linewidth]{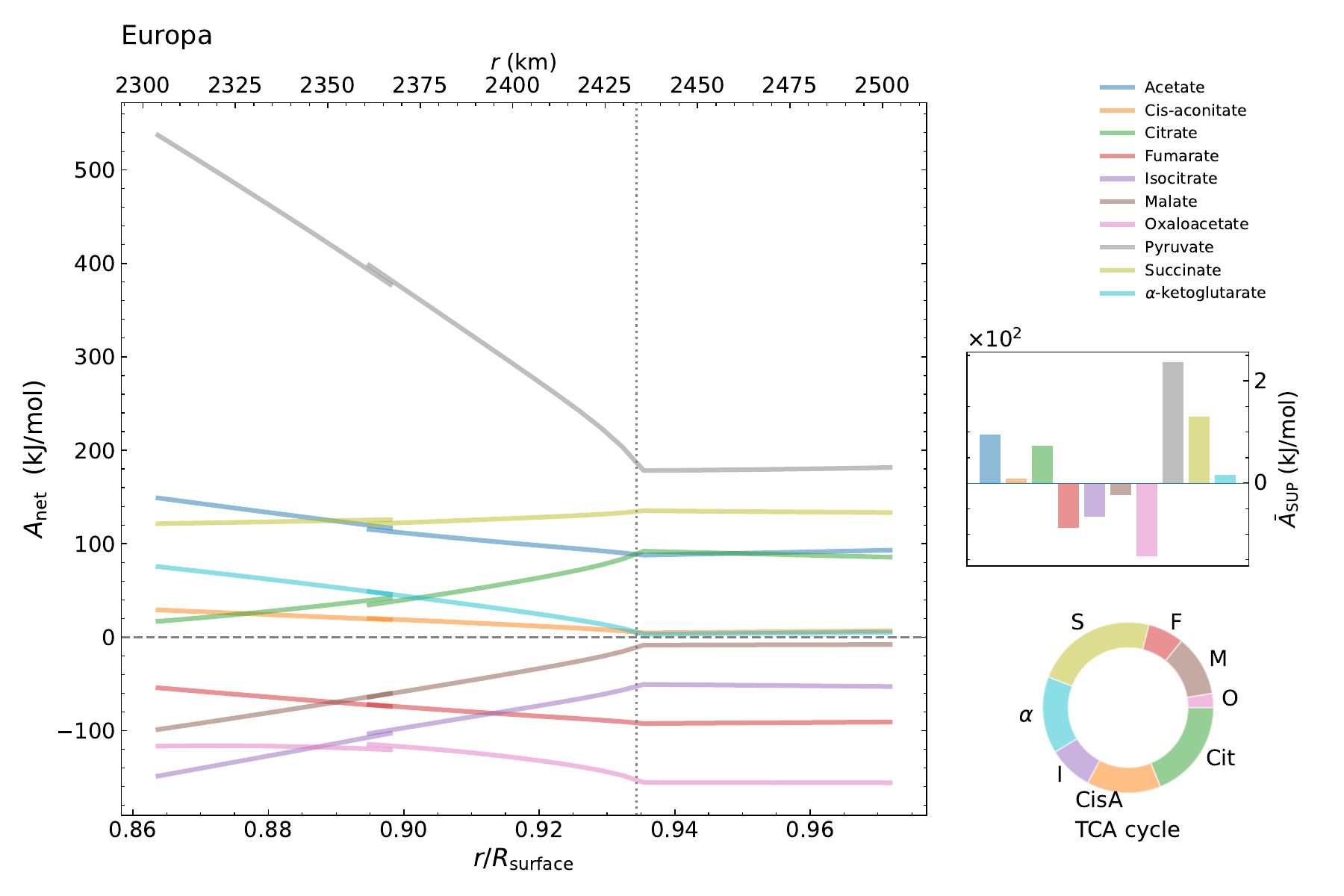}
    \caption{Radial profiles of the net {affinity} 
    for all species entering the TCA and prebiotic networks 
    in aqueous regions of Europa's interior. The curves on the 
    left-hand and right-hand sides are, respectively, the 
    interpolated values from the DEW and the SUPCRT models. 
    The dashed and dotted lines show the critical radii, above 
    which, respectively the DEW and SUPCRT results were extrapolated. 
    Also shown are the {volume-averages of the affinity} along 
    the SUPCRT curves (bar plot) and the relative {$A_{\rm net}$ 
    distribution of the TCA-cycle species} (annular plot).}
    \label{fig:Europa-summary}
\end{figure*}

\begin{figure*}
    \centering
    \includegraphics[width=0.85\linewidth]{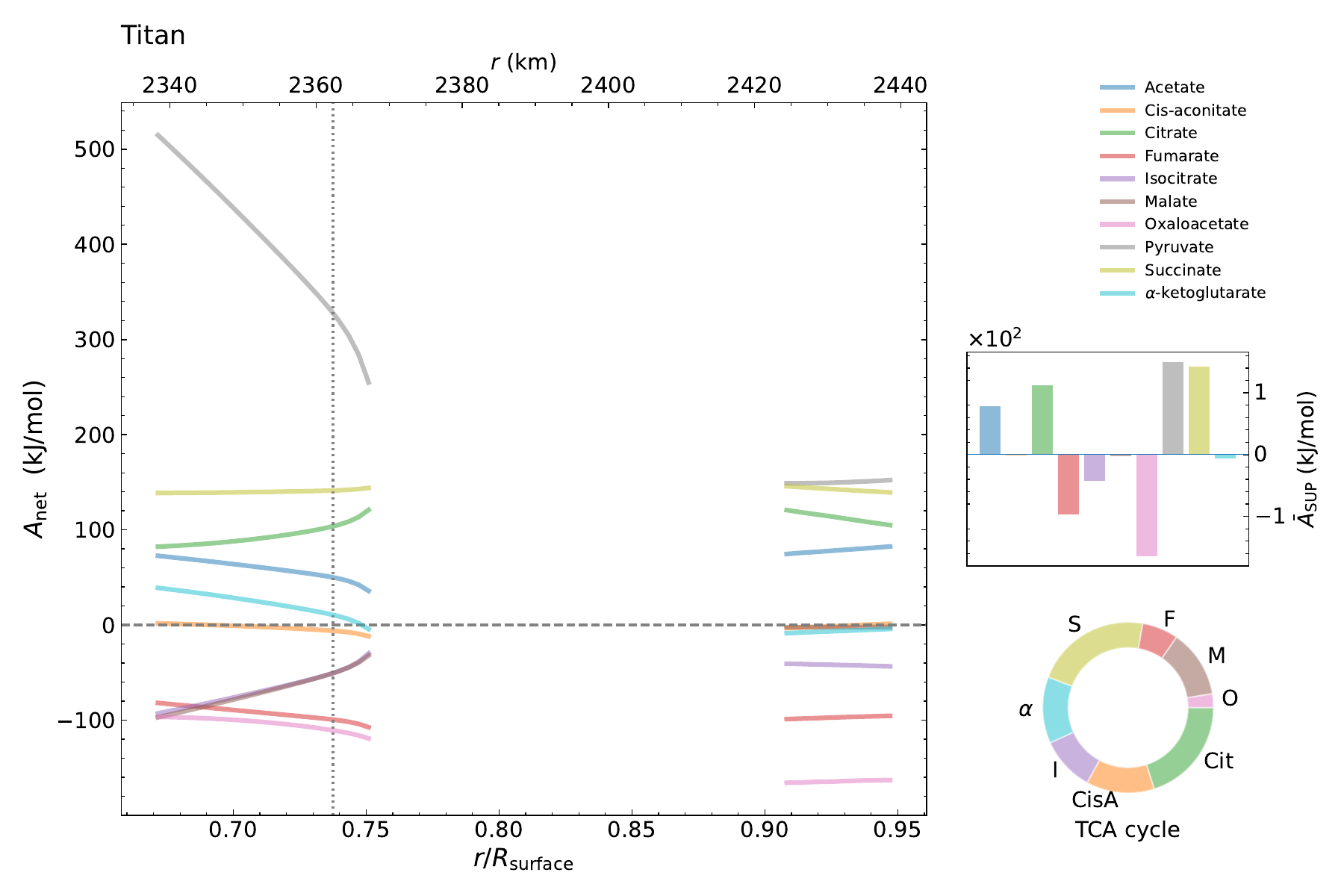}
    \caption{{Radial profiles of the net {affinity} 
    for all species entering the TCA and prebiotic networks 
    in aqueous regions of Titan's aqueous interior layers} (similar to Fig.~\ref{fig:Europa-summary}). 
    The outer-ocean profiles are entirely extrapolated from the 
    SUPCRT model (see Fig.~\ref{fig:TCA}), and the $\sim$5-km part to the right side of the vertical line are extrapolated from the DEW model.}
    \label{fig:Titan-summary}
\end{figure*}

\begin{figure*}
    \centering
    \includegraphics[width=0.85\linewidth]{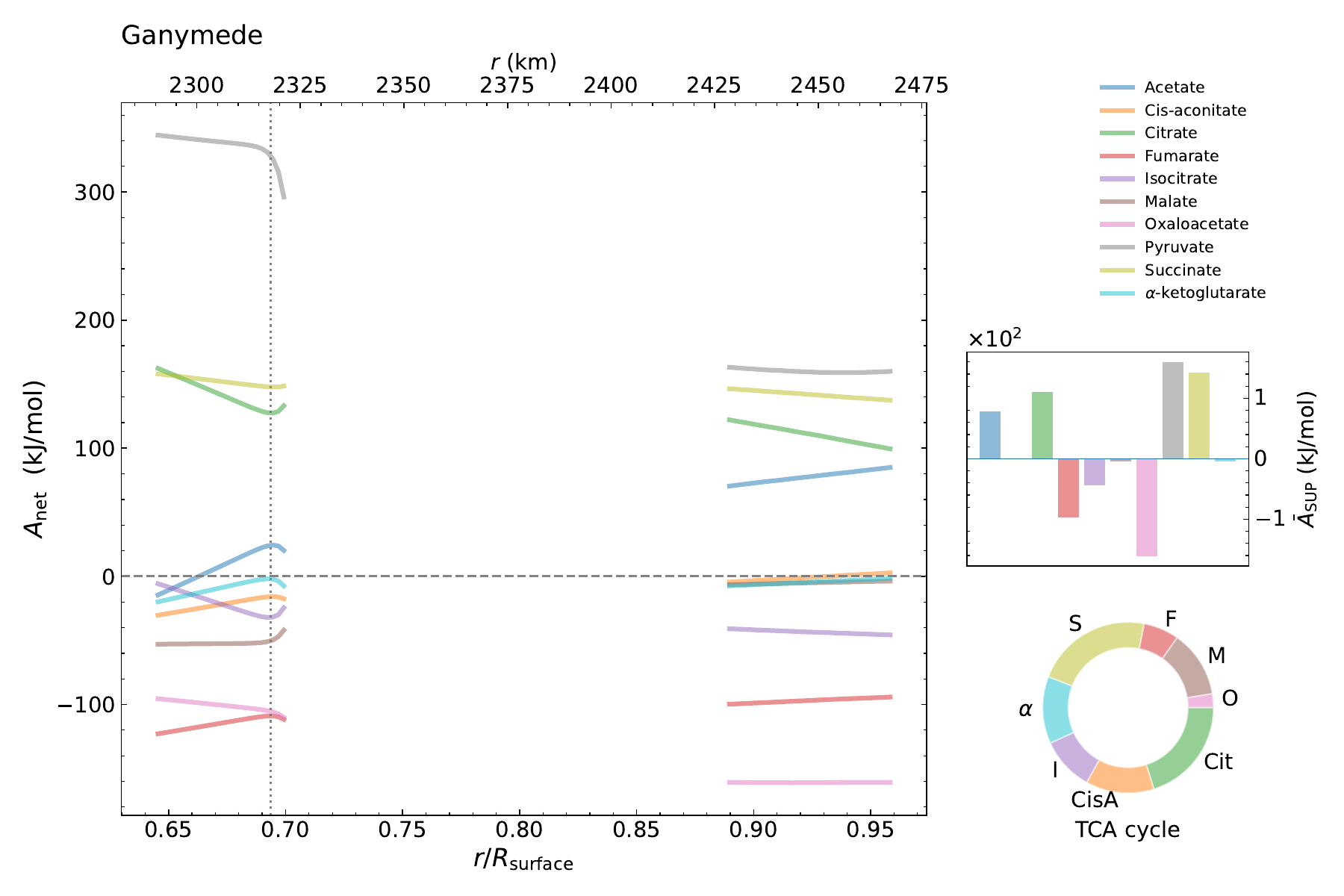}
    \caption{{{Radial profiles of the net affinity
    and their volume averages
    through the aqueous regions of Ganymede's interior regions} (similar to Fig.~\ref{fig:Europa-summary}}.}
    \label{fig:Ganymede-summary}
\end{figure*}

\begin{figure*}
    \centering
    \includegraphics[width=0.85\linewidth]{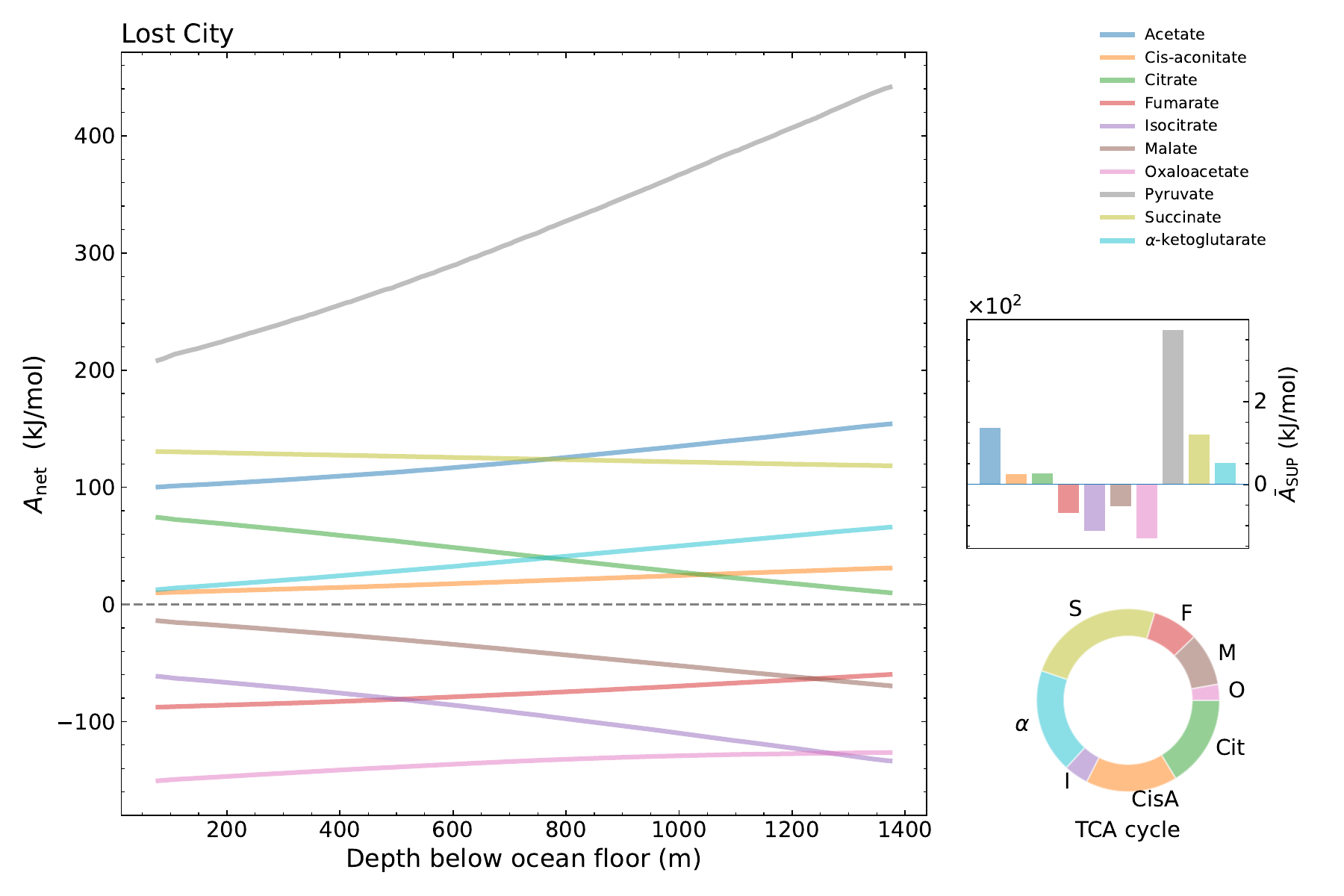}
    \caption{{Radial profiles of the net {affinity} 
    and their volume averages
    through the Lost City hydrothermal field borehole} (similar to Fig.~\ref{fig:Europa-summary}. 
    All points were interpolated along the geotherm using the SUPCRT results (see Fig.~\ref{fig:TCA}).}
    \label{fig:LostCity-summary}
\end{figure*}

We now focus on the net {affinity} for a 
given chemical species, $A_{\rm net}$ (following Eq.~\ref{eq:Anet}; see Sect.~\ref{ssec:stab}). 
This quantity {combines} the {affinities} of those 
reactions which involve the species considered, i.e., each 
node in the reaction network. In the present analysis, however, 
we exclude nodes with only one connection (e.g., glucose), for 
simplicity. 
$A_{\rm net}$ provides a measure for the {net energy production involving reactions in which} the species {is}
involved. {$A_{\rm net}>0$ (or $A_{\rm net}<0$) indicates net production (or consumption) of the node species, when reactions begin with the prescribed non-equilibrium concentrations (e.g., Table~\ref{tab:species}).} In Figs.~\ref{fig:Europa-summary}-\ref{fig:LostCity-summary},
we show the {net affinity} profiles for all those species having more than one edge reaction in Fig.~\ref{fig:network}. 

In Europa (Fig.~\ref{fig:Europa-summary}), the regions covered by SUPCRT and DEW overlap around a fractional radius of $r/r_{\rm surface}=0.90$, because the aqueous parts of the Europan interior model does not have any discontinuity (ie., the ocean sits on top of the rocky interior, which could potentially host liquid water in pores or fractures; see Fig.~\ref{fig:seafreeze}). Both SUPCRT and 
DEW models show good agreement here, which was also evident on the PT plane (Figs.~\ref{fig:res_prebiotic}-\ref{fig:TCA_1}). From this plot, the radial profiles of the 
expected relative proportions of equilibrium abundances can be 
evaluated. Beyond the dotted vertical line ($r \gtrsim 0.935$), 
$A_{\rm net}$ has been extrapolated from the low T 
boundary of the SUPCRT regime. This depth range corresponds to 
the ocean with a near-constant T, which has a stabilizing effect on the radial profiles of $A_{\rm net}$, i.e., all being nearly horizontal. 

As a statistical proxy of the total expected species {stability} under abiotic 
conditions, we also {calculate} the {volume-averaged} $A_{\rm net}$ over 
the SUPCRT radial range, namely, 
\begin{equation}
    \bar{A}_{\rm SUP} = \frac{\int_{\rm SUP} A_{\rm net}r^2dr}{\int_{\rm SUP}r^2dr}
\end{equation}
where the `SUP' range of the integral corresponds to about  
0.895 to 0.975$R_{\rm Surface}$ {for Europa (see below for the respective ranges in the other bodies).} The integral is evaluated over the volume of the spherical shell bounded by the radial limits of the SUPCRT range of calculation. 
For the Lost City site, {we consider}
the distance over which the borehole measurements were made (from 78 m to 1374 m below the seafloor \cite{expedition_340t_scientists_integrated_2012}) in the serpentinized portion ($\sim$180 km$^3$ \cite{fruh-Green+03}) of the dome-shaped Atlantis Massif peridotites. Under the $PT$ conditions of Europa's ocean and the sub-ocean part, {the formation of citrate, cis-aconitate, $\alpha$-ketoglutarate, and succinate are favored, while all the others are more likely to be consumed, when activities from Canovas \& Shock\citep{Canovas+Shock20} are used. However, if we consider individual reaction equilibria only (see $\Delta G$ plots in Figs.~\ref{fig:res_prebiotic}-\ref{fig:TCA_1}), then}
the formation of acetate, 
succinate, pyruvate, malate, and citrate are favored, while 
$\alpha$-ketoglutarate, cis-aconitate, fumarate, and oxaloacetate are 
depleted. {In the non-equilibrium picture, pyruvate is very strongly produced by its adjacent reactions. Cis-aconitate, malate, and $\alpha$-ketoglutarate are relatively near the zero line, i.e., their} positive production 
rate throughout the ocean is nearly balanced by its consumption below the ocean, leading to a relatively small value of {$\bar{A}_{\rm net}$}. 

Also shown in Fig.~\ref{fig:Europa-summary} (lower-right) is a ring chart 
that shows {the volume-averaged} $\bar{A}_{\rm net}$ for the species that enter 
the TCA cycle, {taking into account the connections to the prebiotic network.}
The aerobic biotic TCA cycle would run in the clockwise direction; our prediction for the abiotic TCA cycle in the Europan ocean and its rocky interior reveals an 
alternating pattern for the expected {net affinities using Earth-like activities}: the consecutive 
species in the cycle are {energetically} favored, disfavored, favored, and so on. {The amplitude of $\bar{A}_{\rm net}$ in this species (or reaction) spectrum is thus modulated with a `wavelength' of about two species, over which it rises and falls between positive and negative affinities. This is naturally expected from such a system, where the species are interconnected by adjacent reactions. The total energy  in a closed system is conserved, so one would expect that the sum $\sum_s\bar{A}_{\rm SUP}$ over all species $s$ to be nearly zero. In the present case, however, the TCA cycle is also externally manipulated by connections through acetate and oxaloacetate (Fig.~\ref{fig:network}), hence there will in general be a nonzero sum of the net affinity for an arbitrary non-equilibrium distribution of initial activities, leading to such alternating patterns in net species affinities.}

We now summarize similar plots for the other ocean worlds and the Lost City Hydrothermal Field-adjacent borehole. In both Titan (Fig.~\ref{fig:Titan-summary}) and Ganymede (Fig.~\ref{fig:Ganymede-summary}), there is a wide separation between the two aqueous ranges, as 
there are layers of ice V and VI in between (see Fig.~\ref{fig:seafreeze}). Owing to their high P and low T profiles, the outer oceans of Europa, Titan and Ganymede are 
not covered by the SUPCRT model, though the ocean PT profiles are close to the SUPCRT range. We thus extrapolated the $A_{\rm net}$ values 
for the outer oceans of these three moons (see, e.g., Fig.~\ref{fig:res_prebiotic} and Sect.~\ref{ssec:waterphases}). 

In Titan's ocean and the rocky interior regions where liquid water may be stable, {some species} show steeper radial gradients {than others}. This is related to the uneven contributions of their various 
reactions at different depths (see Sect.~\ref{ssec:stab}). 
All in all, the {volume-averaged net affinity} of the various species shows 
a very similar distribution compared to that of Europa, although the net depletion 
in some species is {more} severe. 

Ganymede (see Fig.~\ref{fig:Titan-summary}) has two isolated ocean layers, 
the deeper one being located between about 0.65 to 0.70$R_{\rm surface}$, 
which is covered by the DEW model. The volume-averaged net affinity shows 
a very similar distribution to the one for Titan. {Although} the total aqueous volume in Ganymede's outer ocean is larger {than in Titan, the different regions average out to yield almost the same mean distribution of species}. 

Enceladus has a sub-crustal ocean between about 0.77 and 0.92$R_{\rm surface}$, which is very near the SUPCRT calculation range. Similar to the other moons, the net {affinity} was extrapolated in that domain, though it is the closest ocean to the PT range covered by the SUPCRT model (and partly covered by forcing the model; see Sect.~\ref{ssec:thermo-models}). Overall, the {net affinity averaged} over the aqueous volume of Enceladus looks {very} similar to the distribution at Europa.

{Figure~\ref{fig:LostCity-summary} shows the same distributions} for the Lost City site, {where} the top boundary of the $\sim$1.3-km-deep probe has a pressure slightly higher than twice that of the Enceladus' seafloor. The resulting $\bar{A}_{\rm SUP}$ distribution is thus similar to that in Enceladus integrated over its entire interior, {all of which} is fully encompassed by the SUPCRT range). 
The {reaction} network contains the same bottlenecks, most severely for oxaloacetate and {isocitrate}. 

\subsection{Enceladus case study}
\label{ssec:enceladus}

{Among the moons considered in this study, Enceladus is the most well-known in terms of the relative concentrations of basic inorganic species. Here, we evaluate the effects of observationally inferred and modeled inorganic concentrations on the net affinity for each central species in our network. To do so, we {collect information from the literature and attempt to} constrain the activities of all large organic species {as well as the smaller species} in the network, by approximating them with their equilibrium concentrations in formation reactions. }
{
The comparison between terrestrial (E. coli) and Enceladus-specific activities provides an important test of our methodology. While the dominant stability patterns (e.g., oxaloacetate as a bottleneck species) persist across both scenarios, the quantitative differences highlight how environmental conditions can modify reaction favorability. This suggests our predictions are robust to different activity assumptions while remaining sensitive to realistic chemical variations between environments.
}

\subsubsection{Estimating {equilibrium} activities of organics}
\label{ssec:formation}
{To estimate the activities of the various organics involved in the network, we model reactions that lead to their formations out of two basic compounds, $\ce{CO2}$ and $\ce{H2}$, requiring stoichiometric and charge balance. The reactions are listed in Table~\ref{tab:formation}. Using DEWPython, we first determine the equilibrium constant and the Gibbs free energy change for these reactions in the respective PT range of SUPCRT. The species are largely favored in the equilibrium. }


\begin{table*}
    \centering
    \begin{tabular}{ll}
    \hline\hline 
    Citrate & $\ce{6CO2} + \ce{9H2} \ce{<=>} \ce{citrate^{3-}} + \ce{5H2O} + \ce{3H^{+}}$\\
    \hline 
    Cis-aconitate & $\ce{6CO2} + \ce{9H2} \ce{<=>} \ce{cis-aconitate^{3-}} + \ce{6H2O} + \ce{3H^{+}}$\\
    \hline 
    Isocitrate & $\ce{6CO2} + \ce{9H2} \ce{<=>} \ce{isocitrate^{3-}} + \ce{5H2O} + \ce{3H^{+}}$\\
    \hline 
    $\alpha$-ketoglutarate & $\ce{5CO2} + \ce{8H2} \ce{<=>} \ce{\alpha{\rm -ketoglutarate^{2-}}} + \ce{5H2O} + \ce{2H^{+}}$\\
    \hline 
    Succinate & $\ce{4CO2} + \ce{7H2} \ce{<=>} \ce{succinate^{2-}} + \ce{4H2O} + \ce{2H^{+}}$\\
    \hline 
    Fumarate & $\ce{4CO2} + \ce{6H2} \ce{<=>} \ce{fumarate^{2-}} + \ce{4H2O} + \ce{2H^{+}}$\\
    \hline 
    Malate & $\ce{4CO2} + \ce{6H2} \ce{<=>} \ce{malate^{2-}} + \ce{3H2O} + \ce{2H^{+}}$\\
    \hline 
    Oxaloacetate & $\ce{4CO2} + \ce{5H2} \ce{<=>} \ce{oxaloacetate^{3-}} + \ce{3H2O} + \ce{2H^{+}}$\\
    \hline 
    Alanine & $\ce{3CO2} + \ce{NH4} + \ce{6H2} \ce{<=>} \ce{alanine^{-}} + \ce{4H2O} + \ce{H^{+}}$\\
    \hline 
    Pyruvate & $\ce{3CO2} + \ce{5H2} \ce{<=>} \ce{pyruvate^{-}} + \ce{3H2O} + \ce{H^{+}}$\\
    \hline  
    Glucose & $\ce{6CO2} + \ce{12H2} \ce{<=>} \ce{glucose} + \ce{6H2O}$\\
    \hline  
    Lactate & $\ce{3CO2} + \ce{6H2} \ce{<=>} \ce{lactate^{-}} + \ce{3H2O} + \ce{H^{+}}$\\
    \hline  
    Acetate & $\ce{2CO2} + \ce{4H2} \ce{<=>} \ce{acetate^{-}} + \ce{2H2O} + \ce{H^{+}}$\\
    \hline 
    \end{tabular}
    \caption{Formation reactions}
    \label{tab:formation}
\end{table*}

Next, we consider the estimated activities of $\ce{CO2}$, $\ce{H2}$, and $\ce{H+}$ as equilibrium concentrations, by adopting values representative of the Enceladean ocean from Waite et al.\citep{Waite17} for $pH=9$, assuming the activity of water to be unity. To estimate the corresponding equilibrium concentrations of the organics along the interior of Enceladus, we first interpolate the $\log K$ values on the $PT$ plane at locations of the $PT$ curve of the Enceladean interior (calculated by PlanetProfile, see Sect.~\ref{ssec:waterphases}). We then consider the case of zero affinity and solve the equation $\log Q = \log K$ for the unknown activity $\log a_s$ of a given species $s$, as a function of the radial location. {This provides concentrations corresponding to formation equilibria.} In general, the inferred concentrations increase with height, except for the ocean, where they keep a rather constant level, which we take as the final activity. 

{It turns out that in most cases the equilibrium concentration increases from lower values in the deep interior to unreasonably high values in the ocean, beyond solubility limits. Remarkably, all the equilibrium concentrations end up at {such high} levels, except for oxaloacetate, for which the formation from inorganics leads to a concentration that is several orders of magnitude lower, $5\times 10^{-8}$~\si{kg/mol}. 
The {cofactor} species entering the reaction network, {namely NAD$^-_{\rm ox}$, NAD$^{-2}_{\rm red}$, NADH, NAD$^+$, ADP$^{-3}$, and ATP$^{-4}$, are complex species known to exist only on Earth's biosphere to date, though there is no counter-evidence for their formation in abiotic aqueous environments. To be conservative,} we assume an arbitrarily low concentration of $10^{-12}$~\si{kg/mol} to each of them, also because no information as to their possible values in Enceladus was found in the literature. Finally, for $\ce{PO4^{-3}}$, \ce{NH3} and \ce{HCO3-} we use reported concentrations of $\ce{HPO4^{-2}}$\cite{postberg_detection_2023}, \ce{NH3}\cite{Waite17} and \ce{CO2}\cite{Waite17}, along with the equilibrium constants of the respective formation reactions using DEWPython, {leading to their activity estimates}: }
\begin{eqnarray}
    \ce{HPO4^{-2}} &\ce{<=>}& \ce{PO4^{-3}} + \ce{H+} \\
    \ce{NH4} &\ce{<=>}& \ce{NH3^{-}} + \ce{H+} \\
    \ce{HCO3^{-}} &\ce{<=>}& \ce{CO2} + \ce{OH-}.
\end{eqnarray}
{The adopted concentrations are given in Table~\ref{tab:species}, sorted by descenting molality.}

\begin{table*}
    \centering
    \begin{tabular}{llll}
    \hline
    Species & Activity (kg/mol) & Reference \\
    \hline 
    \ce{HPO4}$^{-2}$& $1\times 10^{-3}$ & \ {Postberg et al.\citep{postberg_detection_2023} (2023)}\\
    \ce{HCO3}$^{-}$& $1\times 10^{-3}$ & \ {Waite et al.\citep{Waite17} (2017); present study}\\
    \ce{H2}& $1\times 10^{-4}$ & \ {Waite et al.\citep{Waite17} (2017)}\\
    \ce{CO2}& $7\times 10^{-5}$ & \ {Waite et al.\citep{Waite17} (2017)}\\
    \ce{NH3}& $6\times 10^{-5}$ & \ {Waite et al.\citep{Waite17} (2017); present study}\\
    \ce{OH}$^{-}$ & $1\times 10^{-5}$ & \ {Glein et al.\citep{GLEIN_2015} (2015)}\\
    \ce{PO4}$^{-3}$& $3\times 10^{-7}$ & \ {present study}\\
    Oxaloacetate$^{-2}$& $5\times 10^{-8}$ & \ {present study}\\
    \ce{H+}& $1\times 10^{-9}$ & \ {Glein et al.\citep{GLEIN_2015} (2015)}\\
    \hline
    \end{tabular}
    \caption{Estimated activities of species for Enceladus}
    \label{tab:species}
\end{table*}

{For the remaining organics with large molecular weight (i.e., the node species of the entire network in Fig.~\ref{fig:network}, for which the equilibrium concentrations estimated above are unreasonably high), we estimate aqueous concentrations using data from studies of carbonaceous chondrites investigated in meteorites in the next section\citep{Pizzarello2006,Glavin2010}. }

{\subsubsection{Estimating the activities of larger soluble organics}}
\label{sssec:som}
{The concentrations of organic molecules in carbonaceous chondrites are often reported in parts per million (ppm), which is a mass-based unit (e.g., \si{\micro\gram} of compound per \si{\gram} of meteorite). To use these measurements as proxies for the expected organic concentrations in water, a series of conversions is in order.}

{We first convert the concentration of each organic compound in \si{ppm} to \si{nmol/g} using the molecular weight (\si{g/mol}) of the compound:
\begin{equation}
    C_{\si{nmol/g}} = \frac{C_{\si{ppm}} \times 1000}{M},
\end{equation}
where \( C_{\si{nmol/g}} \) is the concentration in \si{nmol/g}, \( C_{\si{ppm}} \) is the concentration in \si{ppm}, $M$ is the molar mass of the compound in \si{g/mol}.}

{To estimate the aqueous concentrations in oceans, we consider dilution of the meteoritic concentrations by a large volume of water. Assuming a water-to-rock ratio $W/R$, we calculate the aqueous concentration (\( C_{\si{\micro M}} \)) using 
\begin{equation}
    C_{\si{\micro M}} = \frac{C_{\si{nmol/g}}}{W/R}.
\end{equation}
We finally convert the aqueous concentrations in \si{\micro M} to \si{kg/mol}, using the molar mass of the compound in \si{kg/mol}:
\begin{equation}
    C_{\si{kg/mol}} = C_{\si{\micro M}} \times 10^{-6} \times M_{\si{kg/mol}}.
    \label{eq:som}
\end{equation}
The resulting concentrations are given in Table~\ref{tab:som}, along with the compound class, the range of meteoritic concentration\citep{Pizzarello2006,Glavin2010}, the inferred mean solution concentration using Eq.~(\ref{eq:som}), and remarks regarding the solubility and stability of the species. The entries are sorted by descending molality. Intriguingly, the inferred aqueous concentration of oxaloacetate is close to the value we estimated from formation equilibrium (see Table~\ref{tab:species}). This indicates that, although oxaloacetate becomes unstable when dissolved within water, the inferred low concentrations are quantitatively consistent with its thermodynamic equilibrium with $\ce{CO2}$ and $\ce{H2}$ in the Enceladean ocean.}

\begin{table*}[h!]
\centering
\caption{Organic concentrations in meteorites and aqueous environments}
\begin{tabular}{lllllll}
\toprule
Compound & $M$ & Class & Meteoritic & $\langle C_{\rm aq}\rangle^{1}$ & Notes \\
    & (\si{g/mol})  &  & $C_{\rm ppm}$\citep{Pizzarello2006,Glavin2010} & ($10^{-7}$~\si{kg/mol})  &   \\
\midrule
Acetate$^{-}$           & 59.04                   & Carboxylic     & 50–500                                      & \num{158}                &                   Highly soluble \\
Pyruvate$^{-}$          & 87.06                   & Keto acid      & 10–50                                       & \num{11.7}                                    & Moderately soluble \\
Alanine           & 89.09                   & Amino acid     & 10–50                                       & \num{11.4}                &                   Relatively soluble \\
Lactate           & 90.08                   & Carboxylic     & 10–50                                       & \num{11.3}                &                   Moderately soluble \\
Fumarate$^{2-}$          & 116.07                  & Dicarboxylic   & 10–50                                       & \num{8.77}                &                   Moderately soluble \\
Succinate$^{2-}$         & 118.09                  & Dicarboxylic   & 10–50                                       & \num{8.62}                &                   Moderately soluble \\
Malate$^{2-}$            & 134.09                  & Dicarboxylic   & 10–50                                       & \num{7.59}                                    & Moderately soluble \\
$\alpha$-Ketoglutarate$^{2-}$ & 146.10          & Keto acid      & 10–50                                       & \num{6.97}                 &                  Moderately soluble \\
Citrate$^{3-}$           & 189.10                  & Tricarboxylic  & 10–50                                       & \num{5.39}                & Moderately soluble \\
Oxaloacetate$^{2-}$      & 132.07                  & Dicarboxylic   & 1–5                                         & \num{0.771}                &                   Unstable in solution \\
Cis-Aconitate$^{3-}$     & 173.10                  & Tricarboxylic  & 1–5                                         & \num{0.588}                & Less stable \\
Glucose           & 180.16                  & Sugar          & 1–5                                         & \num{0.565}                &                   Rare; low solubility \\
Isocitrate$^{3-}$        & 192.12                  & Tricarboxylic  & 1–5                                         & \num{0.530}                &                  Low solubility, less stable \\
\bottomrule
\end{tabular}
\begin{flushleft}
\footnotesize
$^1$ The average of the minimum and maximum estimates.
\end{flushleft}
\label{tab:som}
\end{table*}

\begin{figure*}
    \centering
    \includegraphics[width=.6\linewidth]{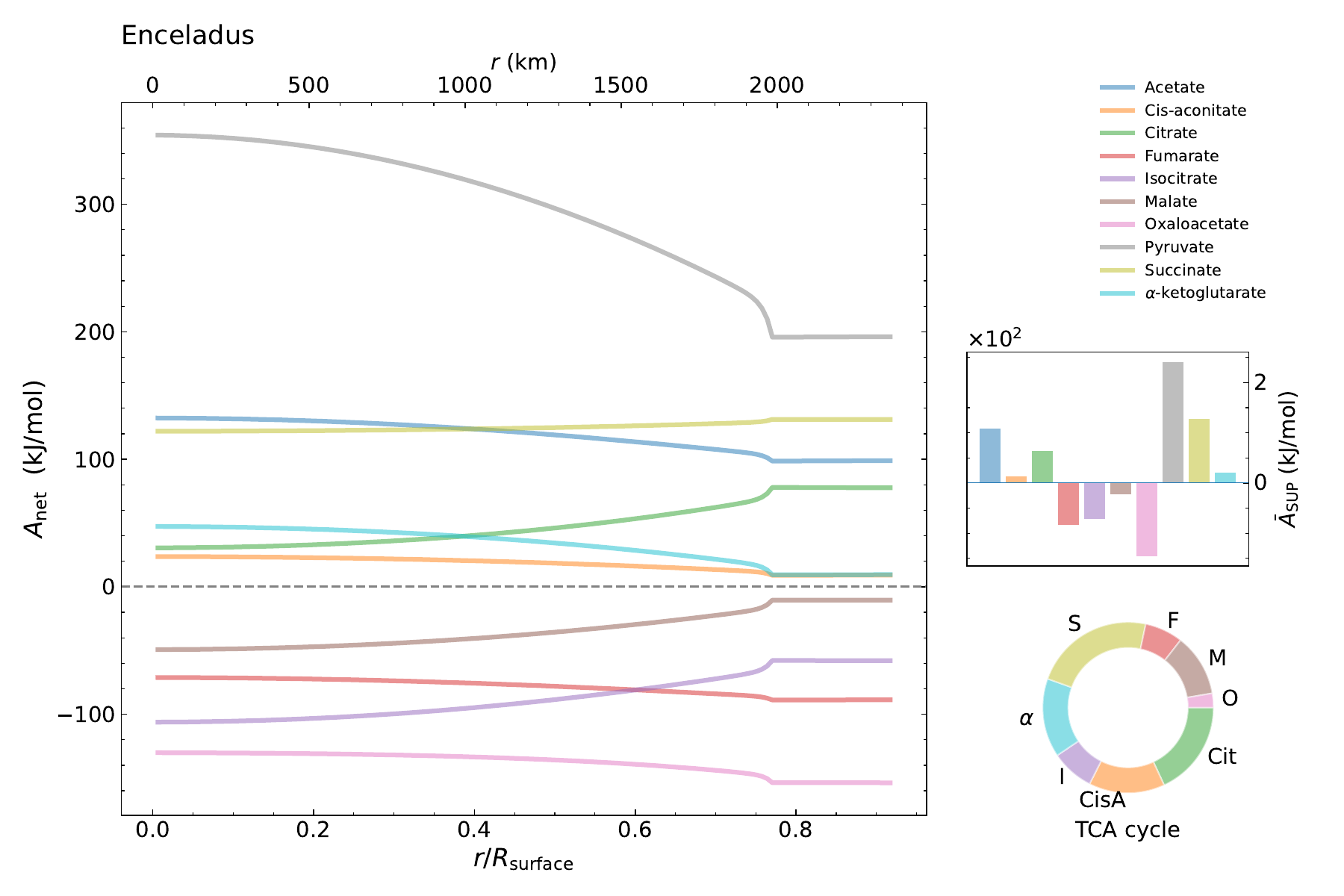}    \includegraphics[width=.6\linewidth]{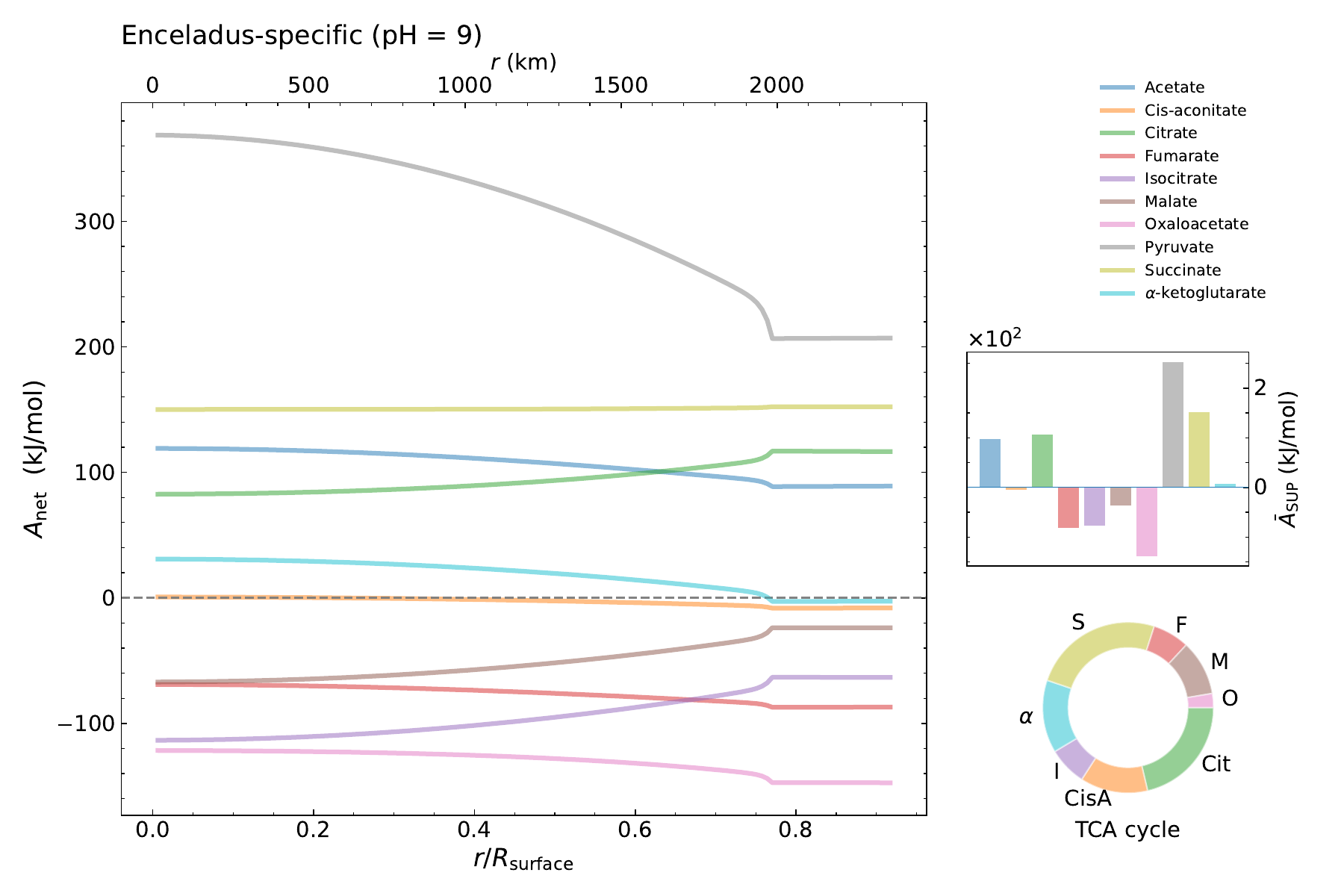}
    \includegraphics[width=.6\linewidth]{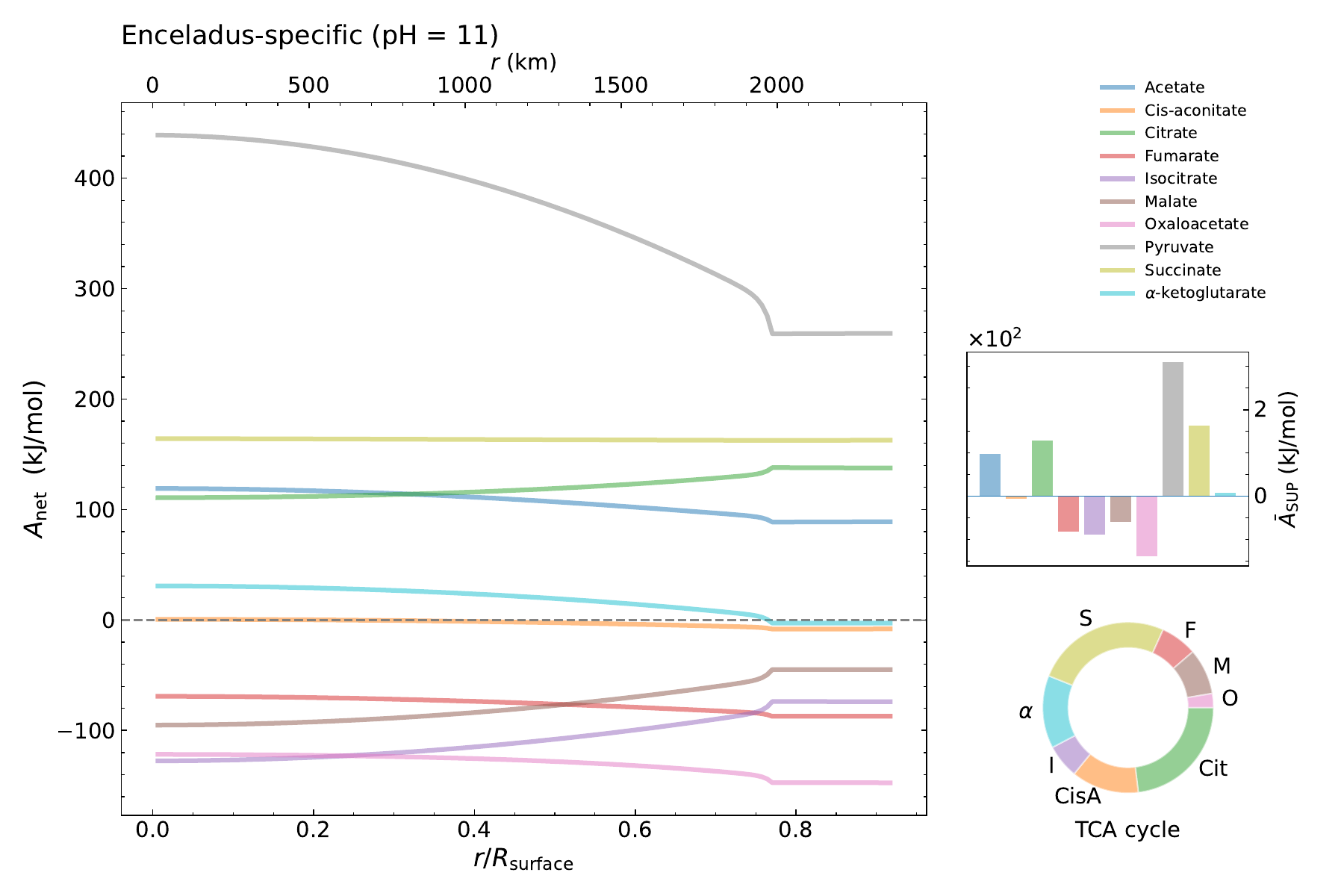}
    \caption{{Radial profiles of the net {affinity} 
    and their volume averages
    through Enceladean ocean (above 0.75$R_{\rm surface}$), using E. Coli-based activities from Canovas \& Shock\cite{Canovas+Shock20} (top panel), and Enceladus-specific activities from Table~\ref{tab:formation} with pH$=9$ (middle panel) and pH$=11$ (bottom panel).}}
    \label{fig:Enceladus-case-summary}
\end{figure*}

{Figure~\ref{fig:Enceladus-case-summary} shows the net affinities comparatively, {assuming E. Coli activities\cite{Canovas+Shock20}, and assuming Enceladus-specific activities for the lower and upper limits of the estimated pH range, 9 and 11.} A number of similarities and differences stand out in the two cases. The dominantly stable and unstable species are unaffected: oxaloacetate, fumarate and isocitrate as the strongly unstable, {pyruvate}, citrate, succinate, and acetate as the strongly stable species. The main differences turn out to be for those species that show near-equilibrium net affinities (e.g., cis-aconitate {and $\alpha-$ketoglutarate}), which are more sensitive to changes in the relatively small absolute affinities, owing to the different sets of activities chosen for all the species. {All in all, the resulting net affinity per species is not very sensitive on the details of the assumed activity distribution, but is largely determined by the equilibrium features (the Gibbs free energy change) associated with each reaction.}

We note that acetate is the only common species with the recent study Liu et al. (2024)\cite{Liu2024}. We find affinity values comparable to their findings for a pH of 9, when the formation reaction of acetate along the Enceladus PT curve is considered. }

\section{Discussion}
\label{sec:discuss}


The chemical energy released from abiotic geochemical reactions likely played a critical role in the emergence and establishment of life on Earth, particularly in environments with water-rock interactions (e.g. Shock and Boyd \cite{shock_principles_2015}). {Our inclusion of biological cofactors (ATP, NAD) in the reaction network requires careful interpretation. While these molecules are typically associated with evolved metabolic processes, we consider only their thermodynamic roles in the overall reaction stoichiometry, not their sophisticated biological functions as energy currencies or electron carriers. This allows us to evaluate whether these reaction sequences could proceed spontaneously under abiotic conditions, even if the specific mechanisms and intermediates differed from modern biochemistry.} 

{Recent experimental work supports the thermodynamic feasibility of our approach within realistic geochemical contexts. Clay minerals and hydrothermal vent structures can adsorb and concentrate organic compounds, substantially enhancing reaction efficiency.\cite{Dalai2020} Additionally, Fe-S peptides have been shown to generate pH gradients across proto-cell membranes that can drive ATP synthesis \cite{Bonfio2018}, providing a mechanism for maintaining non-equilibrium conditions favorable to our modeled reactions. These findings align with the hypothesis that the precursors of modern metabolism emerged in localized geochemical niches where thermodynamic barriers could be overcome through mineral catalysis, compartmentalization, and energy coupling to ambient gradients.}

Here, we have shown the predictive {potential} of modeling reaction {(non-)}equilibria (see also \cite{Krissansen-Totton2016} for chemical disequilibria in atmospheres or atmosphere/ocean systems). We have placed constraints on the feasibility of the TCA cycle and a plausible prebiotic network leading to the TCA cycle. 
In all the sites studied, including the inhabited Lost City Hydrothermal Field, energetic barriers cause bottlenecks in the TCA cycle, preventing a spontaneous reaction cycle (i.e., precluding perpetual energy). The energy barriers we found for the production of species like oxaloacetate, fumarate, and isocitrate may be such that it would be possible for cells to run these reactions endergonically in the forward direction by supplying energy from metabolic resources, so that the TCA cycle can work in the forward direction. The same is true for the reverse TCA cycle, where energy should be provided to reverse reactions such as oxaloacetate to citrate (see Fig.~\ref{fig:network}). 

Our results could aid the interpretation of compositional data to be obtained by space missions such as ESA's JUICE, visiting Europa and Ganymede, NASA's Europa Clipper, NASA's mission to Titan Dragonfly, and possible future missions to Enceladus and dwarf planet Ceres, which are targets prioritized by space agencies and the National Academy of Sciences, Engineering and Medicine \cite{national_academies_of_sciences_origins_2023, martins_report_2024}. Specifically, if the instruments on these missions are able to measure the species involved in the TCA cycle, significant deviations of measured relative concentrations from the equilibrium {and non-equilibrium} chemistry predictions could be indicative of {other} non-equilibrium {chemical states} (i.e., deviation from abiotic equilibrium), which is one of the indications for metabolism, a high rung in the Ladder of Life Detection\cite{neveu_ladder_2018}. Detected species distributions more like the ones in Sect.~\ref{ssec:species}, however, would likely imply the opposite, namely an abiotic soup of complex chemistry, in equilibrium {or alternative non-equilibrium states}. We note that such implications would require treating more complex reaction networks, supported by upcoming experimental studies. Machine learning (ML) models {specifically developed to learn and predict reaction networks\citep{stocker+20,wen+23} can also help effectively expanding the space of prescribed reactions.} 

{Our thermodynamic analysis does not address the kinetic stability of these compounds, which would be an important consideration for their accumulation in real ocean world environments.}
Other processes such as sluggish or rapid kinetics, or fractionation (e.g., from the aqueous phase where an organic molecule might reside, to plumes that could be more directly measured by spacecraft) could alter the distribution of species present. {Assembly theory and network topology approaches applied on chemical species of different complexity and the associated reaction networks are also promising toward constructing alternative scenarios of abiogenesis.\cite{Marshall2021,Sharma2023,Fisher2023}}.

To {evaluate the availability of organics in our network reactions}, we computed the equilibrium constants for possible reactions that would form the species of the TCA cycle from two basic geochemical compounds that could exist in abundance, namely \ce{H2} and \ce{CO2} {(see Sect.~\ref{ssec:formation})}, similar to Amend et al. \cite{amend+11}. The formation of TCA species, including the bottleneck species oxaloacetate and fumarate, is very strongly favored along the PT profiles of the ocean worlds and Lost City. Provided that there is sufficient dissolved $\ce{CO2}$ and $\ce{H2}$ as feedstock, the depletion of some TCA species would not pose a problem for their availability {(see Table~\ref{tab:formation})}. Therefore, theoretical estimation and future measurements of inorganic building blocks of the organic species are essential to further constrain the viability of biochemical reaction networks in deep habitats \citep{Canovas+Shock20}. Observations from the Cassini mission at Enceladus' plumes strongly suggest the presence of \ce{CO2} and \ce{H2} in the ocean \cite{Waite17}, and theoretical predictions, together with observations from the James Webb Space Telescope, Galileo and Juno spacecrafts suggest the presence of \ce{CO2} at Ganymede and Europa \cite{melwani_daswani_metamorphic_2021, trumbo_distribution_2023, villanueva_endogenous_2023, hibbitts_carbon_2003, hansen_widespread_2008}.

We have shown that the small network of prebiotic reactions proceeding from glycolysis to the formation of citrate through acetate can help consistent forward-TCA cycles even in the absence of organisms. {In the analysed case, this occurs in such a way that bypasses the bottleneck species oxaloacetate, providing spontaneous energy production by the pyruvate-acetate path.} If that is true amid the complex chemistry and large-scale flows that mix various parts of ocean worlds, then such abiotic self-{regulation} of reaction networks can {assist} formation of more complex macromolecules, potentially leading to life. Furthermore, metabolisms can exploit the energy that is released in several pathways, e.g., the abiotic forward TCA cycle, to reduce entropy and increase molecular and structural complexity. 

{Intriguingly, the net affinity measure defined in this study is robust to changes in the assumed activities of the constituent species of the reactions in the network, as we have shown for the case of Enceladus (Sect.~\ref{ssec:enceladus}). This insensitivity makes it a potentially useful quantity estimating species that a given reaction network would tend to favor. 
 However, we note also that this quantity is based on instantaneous affinities of reactions surrounding a given species, so the changes in species concentrations will be nonlinearly coupled as reactions progress, requiring more complex approaches beyond the scope of the current study.}

Considering the experimental state of the art, it is evident from our analysis that the SUPCRT model does not cover the 
ocean PT ranges of Europa, Ganymede, and Titan, and only very near the Enceladean ocean. {The limitations of current thermodynamic models for low-temperature conditions relevant to ocean worlds represent a key challenge. While our extrapolations appear reasonable based on observed trends and validation against forced sub-273 K calculations (Supporting Information), future experimental work is needed to directly measure thermodynamic properties in these regimes. This is particularly important for reactions involving charged species, where changes in water's dielectric properties at low temperatures could significantly affect equilibria}. Pitzer-based approaches such as the FREZCHEM code \cite{marion_frezchem:_2010} extend to low T, but have not been extended to high P, and are as yet to include adequate organic thermodynamics (e.g., Naseem et al. \cite{naseem_salt_2023}). This would require experiments in low T, high P range. Another experimental need {(independent of the specific reaction network analysed in the present study)} is to measure thermodynamic data relevant to equilibrium calculations of many aqueous species (numerous soluble organics) and mineral groups (e.g. hydrated salts and green rust, but also insoluble organics), which are presently not available in the databases used by SUPCRT and DEW. We anticipate it will be important to add the information obtained from experimental studies to our library in the future, {to help extending the computational work on organic networks to a much larger reaction space. In addition to} laboratory measurements, ML methods can be employed in the estimation of thermodynamic data for those species that have no measured thermodynamic data in the PT ranges covered by SUPCRT and DEW. 

The analysis we proposed in this paper can be applied to interior models of 
water worlds (e.g., super Earths and mini-Neptunes \cite{Glein2024}), as well as exomoons 
with various possible structures allowing liquid water in their warmer 
interiors.

\section{Conclusions}
The search for life on ocean worlds is a multi-faceted journey through various space missions, experiments, and models. In our study, we have used the predictive power of thermodynamic modeling of the stability and energetics of the citric acid cycle and a precursor network under abiotic conditions relevant to the aqueous parts of ocean worlds targeted by future missions. Specifically, we made use of state-of-the-art interior structure models of ocean worlds \citep{styczinski_planetprofile_2023} and accurate thermodynamic models for phases of water \citep{journaux_large_2020}, and we demonstrated a framework for calculating radial profiles along a planetary body for the relative abundances of aqueous organic species. Our main conclusions are listed below. 

\begin{itemize}
    \item Through much of the parts of ocean world interiors where liquid water could be present, the TCA cycle shows {quasi-}periodic alternations between forward and backward reaction paths, revealing potential energetic barriers in the cycle. The radial profiles of the relative {net reaction affinities} of various species {indicate} accumulation of certain species (e.g., pyruvate, citrate), and depletion of others (e.g., oxaloacetate, fumarate). 
    \item While the TCA cycle species can be produced solely by dissolved $\ce{CO2}$ and $\ce{H2}$, which are thought to be present in some ocean worlds, the potential barriers posed by the bottleneck species (e.g., oxaloacetate, fumarate) require external energy sources (e.g., cell metabolism, or enhanced kinetics), or alternative pathways of production for the TCA cycle to run consistently in the forward or reverse directions. 
    \item {The net affinity proxy for species stability indicates stable and unstable species in the context of considered reactions. The sensitivity of this proxy on the instantaneous activities of the species involved is remarkably low, providing a robust measure.}
    \item Reaction pathways involving species thought to be essential prebiotic ingredients of life (glucose, pyruvate) can help drive the TCA cycle in all the considered ocean-world interior conditions. For the TCA cycle to run consistently in an abiotic environment, alternative pathways such as oxaloacetate \ce{->} pyruvate \ce{->} acetate \ce{->} citrate (See Fig.~\ref{fig:network}) can bypass the said energy barriers. This finding supports previous results hinting at rich organic chemistry that could potentially support microbial life in the oceans of Europa, Ganymede, Enceladus, and Titan, as well as deep habitats in Earth. 
    \item Future missions measuring species of the TCA cycle and the prebiotic network at ocean worlds will {play a critical role in} explaining any observations that {is consistent with or} deviate strongly from the {chemical reaction affinities} presented here.
\end{itemize}

In future, we plan to apply the present modeling approach to more complex biogeochemical reaction networks extending beyond the TCA cycle, and which may act as other sources and sinks to species in the cycle. Targeting other ocean worlds such as Callisto, exoplanetary oceans \cite{Glein2024} and (potentially) Mimas \cite{rhoden_evolution_2024} would improve our understanding of the habitability of ocean worlds on a wider range of geophysical settings. In this regard, we emphasize the needs for extending the SUPCRT and DEW models (and their water thermodynamic properties) to sub-freezing temperatures (to include salty sub-crustal oceans of the solar system). In addition, determination of the missing thermodynamic data for important soluble and insoluble organic species and minerals will help modeling equilibrium conditions of extensive biogeochemical networks in the solar system's deep water habitats. Similarly, considering that the rise and evolution of life was mediated by redox processes involving iron compounds such as iron sulfide clusters \citep{Beiner97} or \ce{Fe^2+}/\ce{Fe^3+} phases \cite{barge_redox_2019, weber_testing_2022}, we plan to account for such minerals in the reaction networks as well.

\begin{acknowledgement}

S.I. acknowledges funding from the Council of Science and Technology of 
T\"urkiye (T\"UBITAK), under grant 122F287. 
M.M.D. thanks Caltech-JPL summer intern Andrew Chan for helping produce the first iteration of the DEWPython program. M.M.D. was supported by the NASA Planetary Science Early Career Award Program NNH19ZDA001N-ECA to proposal \#19-ECA19-0032.
{Part of the funding for this work was through NASA’s Astrobiology Program’s CAN-8 Project ``Habitability of Hydrocarbon Worlds: Titan and Beyond'', PI Dr. Rosaly Lopes (JPL/Caltech).} A part of this research was carried out at the Jet Propulsion Laboratory, California Institute of Technology, under a contract with the National Aeronautics and Space Administration (80NM0018D0004). \copyright{} 2024. Jet Propulsion Laboratory, California Institute of Technology. Government sponsorship acknowledged. 

\end{acknowledgement}


\begin{suppinfo}
Supporting Information: 1) PT profile of IODP borehole U1309D near Lost City Hydrothermal Field (CSV); 2) Figure showing SUPCRT extrapolated below freezing, plus figures for $\log_{10} K$ and $\Delta G$ of prebiotic network and TCA cycle species formed from aqueous \ce{CO2} and \ce{H2} (PDF). 
{3) E. Coli-based activities of species used in Sect.~\ref{ssec:Ecoli}.}
\end{suppinfo}




\end{document}